\documentclass[10pt,doublecolumn,twoside]{IEEEtran}
\usepackage{cite}
\usepackage{amsmath}
\usepackage{amssymb}
\usepackage{tikz}
\usepackage{pgfplots}
\usepackage{graphicx}
\usepackage{caption}
\usepackage{subcaption}
\usepackage{xcolor}
\allowdisplaybreaks

\newcommand{\until}[2]{\, U_{[{#1},{#2}]} \, }
\newcommand{\rs}[1]{\rho^{#1}(\boldsymbol{s},t)}
\newcommand{\rss}[2]{\rho^{#1}(\boldsymbol{s},{#2})}

\newcommand{\re}[1]{{\color{black}#1}}

\newtheorem{definition}{Definition} % definition numbers are dependent on theorem numbers
\newtheorem{theorem}{Theorem} % same for example numbers
\newtheorem{assumption}{Assumption} % same for example numbers
\newtheorem{problem}{Problem} % same for example numbers
\newtheorem{remark}{Remark}
\newtheorem{lemma}{Lemma}
\newtheorem{corollary}{Corollary}
\newtheorem{example}{Example}

\ifCLASSINFOpdf
\else
\fi

\begin{document}
\title{ Barrier Function-based  Collaborative Control of Multiple Robots under Signal Temporal Logic Tasks}

\author{Lars~Lindemann and Dimos~V.~Dimarogonas
\thanks{L. Lindemann (llindem@kth.se) and D. V. Dimarogonas (dimos@kth.se) are with the Division of Decision and Control Systems, School of Electrical Engineering and Computer Science, KTH Royal Institute of Technology, Stockholm, Sweden.}
\thanks{This work was supported in part by the Swedish Research Council (VR), the European Research Council (ERC), the Swedish Foundation for Strategic Research (SSF),  the EU H2020 Co4Robots project, and the Knut and Alice Wallenberg Foundation (KAW).}% <-this % stops a space
%\thanks{Manuscript received month day, 20xx; revised month day, 20xx.}
}

\maketitle

\begin{abstract}
Motivated by the recent interest in cyber-physical and autonomous robotic systems, we study the problem of dynamically coupled multi-agent systems under a set of signal temporal logic tasks. In particular, the satisfaction of each of these signal temporal logic tasks depends on the behavior of a distinct set of agents. Instead of abstracting the agent dynamics and the temporal logic tasks into a discrete domain and solving the problem therein or using optimization-based methods, we derive collaborative feedback control laws.  These control laws are based on a decentralized control barrier function condition that results in discontinuous control laws, as opposed to a centralized  condition resembling the single-agent case. The benefits of our approach are inherent robustness properties typically present in feedback control as well as  satisfaction guarantees for continuous-time multi-agent systems.  More specifically, time-varying control barrier functions  are used that account for the semantics of the signal temporal logic tasks at hand. For a certain fragment of signal temporal logic tasks, we further propose a systematic way to construct such control barrier functions. Finally, we show the efficacy and robustness of our framework in an experiment including a group of three omnidirectional robots.
\end{abstract}


% \fi
\IEEEpeerreviewmaketitle

\section{Introduction}
\label{sec:introduction}

A multi-agent system is a collection of independent agents with individual  actuation, computation, sensing, and decision making capabilites. Compared to single-agents systems, advantages are scalability with respect to task complexity, robustness to agent failure, and better overall performance. Collaborative control of multi-agent systems deals with achieving tasks such as consensus  \cite{ren2005consensus}, formation control \cite{tanner2003stable}, connectivity maintenance \cite{zavlanos2008distributed}, and collision avoidance \cite{mastellone2008formation} (see \cite{mesbahi2010graph} for an overview). A recent trend has been to extend beyond these standard objectives and to consider more complex task specifications by using temporal logics. Towards this goal, both single-agent systems \cite{kloetzer2008fully,fainekos2009temporal,kress2009temporal} as well as multi-agent systems \cite{loizou2004automatic,guo2015multi,kloetzer2010automatic,sahin2017provably,kantaros2019temporal} have been considered by using linear temporal logic (LTL). Most of these works require a discrete abstraction of the agent dynamics to then employ computationally costly graph search methods. Signal temporal logic (STL) \cite{maler2004monitoring}, as opposed to LTL, allows to impose tasks with strict deadlines and offers a closer connection to the agent dynamics  by the introduction of robust semantics \cite{fainekos2009robustness,donze2}, hence offering the benefit of not necessarily relying on an abstraction of the system. Recent control methods for STL tasks then consider discrete-time systems and result, even for single-agent systems, in computationally costly mixed integer linear programs \cite{raman1,liu2017distributed,belta2018formal}. Control approaches for the non-deterministic setup, still in discrete time, have been presented in \cite{farahani2017shrinking}, while learning-based approaches appeared in \cite{aksaray2016q,muniraj2018enforcing}. An initial approach to obtain satisfaction guarantees for continuous-time multi-agent systems under a fragment of STL tasks has been presented in our previous work \cite{lindemann2018decentralized}. Such continuous-time guarantees have also appeared for single-agent systems in \cite{pant2018fly} where, however, a possibly non-convex optimization problem is to be solved.

Verification of safe sets for dynamical systems has been analyzed by the notion of barrier functions, which are also called barrier certificates. The construction of such barrier functions for polynomial systems using sum of squares programming has been presented in \cite{prajna2007framework}. For control systems and based on the notion of barrier functions, control barrier functions have first been presented in \cite{wieland2007constructive} to guarantee the existence of a control law that renders a desired safe set forward invariant. The authors in \cite{ames2017control} present control barrier functions tailord for safe robot navigation, while \cite{wang2017safety} presents decentralized control barrier functions for safe multi-robot navigation. First robustness considerations of control barrier functions have appeared in \cite{xu2015robustness}. Nonsmooth and time-varying control barrier functions have been proposed in \cite{glotfelter2017nonsmooth} and \cite{xu2018constrained}, respectively. A similar work by the authors of \cite{glotfelter2017nonsmooth} recently proposed hybrid nonsmooth control barrier functions \cite{glotfelter2019hybrid}. In case that such control barrier functions can not be found, safety kernels can be calculated. Safety kernels are  subsets of the safe set that can be rendered  invariant by an active set invariance control method \cite{gurriet2018towards}.  Control barrier functions have also been used to control systems under temporal logic tasks. For single-agent systems, our previous work in \cite{lindemann2018control} has established a connection between the semantics of an STL task and time-varying control barrier functions, while \cite{srinivasan2018control} considers finite-time control barrier functions for LTL tasks. Although both \cite{lindemann2018control} and \cite{srinivasan2018control} deal with achieving finite-time attractivity (see \cite{bhat2000finite} for a definition), the underlying problem definitions differ due to the quantitative, in time and space, nature of STL tasks. Furthermore, \cite{srinivasan2018control} provides upper bounds on the time when a region specified by a static control barrier function is reached, while time-varying control barrier functions provide generic freedom to shape the level sets of a control barrier function at each point in time. Following ideas of \cite{lindemann2018control}, we have presented a collaborative feedback control law for multi-agent systems in \cite{lindemann2019decentralized} where distinct sets of agents are considered and each such set is subject to an STL task; \cite{lindemann2019decentralized}  also presents a procedure to construct control barrier functions for fragments of STL tasks. In contrast to \cite{lindemann2019decentralized}, the work in \cite{lindemann2019control} considers multi-agent systems under possibly conflicting local, i.e., individual, tasks, and deals with finding least violating solutions, so that the problem definitions of \cite{lindemann2019decentralized} and \cite{lindemann2019control} are different.

In this paper, we consider dynamically coupled multi-agent systems under a set of STL tasks. The satisfaction of each task depends on a distinct set of agents. With respect to this setup, the contributions of this paper are threefold. Assuming the existence of control barrier functions that account for the semantics of the STL tasks according to \cite{lindemann2018control}, we first present a collaborative feedback control law that guarantees the satisfaction of all STL tasks. This control law is based on a decentralized control barrier function condition. It turns out, as argued in the technical section of this paper, that this control law is discontinuous so that Filippov solutions and nonsmooth analysis have to be considered. Second, we present an optimization-based approach to construct control barrier functions for a fragment of STL tasks. Third, we provide an experiment that shows the efficacy and robustness of the presented framework. Compared to optimization-based techniques, such as the MILP formulation in \cite{raman1}, the motivation for control barrier function-based techniques is to obtain robust feedback control laws that directly provide STL satisfaction guarantees for continuous-time systems. This paper is an extension of \cite{lindemann2019decentralized}. We here additionally present an experiment of a group of three omnidirectional robots, while we also provide important proofs that are not included in \cite{lindemann2019decentralized}. We also motivate in detail why a discontinuous control law is obtained as opposed to the case where a centralized control barrier function condition is used, resembling the single-agent case. We further extend  \cite{lindemann2019decentralized} by constructing control barrier functions that induce a linear instead of an exponential temporal behavior (as explained in detail in the paper). The advantages of this are shorter computation times to construct the control barrier functions as well as practical benefits such as making it less likely to experience input saturations. 

Section \ref{sec:backgound} states preliminaries and the problem formulation, while our proposed problem solution is stated in Sections \ref{sec:strategy} and \ref{sec:construction_rules}. The experiment using three omnidirectional robots is presented in Section \ref{sec:experiments} followed by conclusions in Section \ref{sec:conclusion}.
\section{Preliminaries and Problem Formulation}
\label{sec:backgound}

True and false are $\top$ and $\bot$, while $\mathbb{R}$ and $\mathbb{R}_{\ge0}$ are the set of real and non-negative real numbers; $\mathbb{R}^d$ is the $d$-dimensional real vector space. Scalars and column vectors are depicted as non-bold letters $v$ and bold letters $\boldsymbol{v}$, respectively. \re{The Euclidean, sum, and max norm of $\boldsymbol{v}$ are $\|\boldsymbol{v}\|$, $\|\boldsymbol{v}\|_1$, and $\|\boldsymbol{v}\|_\infty$,}  respectively. Let $\boldsymbol{0}$ be a vector of appropriate size containing only zeros.  An extended class $\mathcal{K}$ function $\alpha:\mathbb{R}\to\mathbb{R}$ is a locally Lipschitz continuous and strictly increasing function with $\alpha(0)=0$. The partial derivatives, here assumed to be row vectors, of a function $\mathfrak{b}(\boldsymbol{v},t):\mathbb{R}^d\times\mathbb{R}_{\ge 0}\to\mathbb{R}$ evaluated at $(\boldsymbol{v}^*,t^*)$ are $\frac{\partial \mathfrak{b}(\boldsymbol{v}^*,t^*)}{\partial \boldsymbol{v}}:=\frac{\partial \mathfrak{b}( \boldsymbol{v},t)}{\partial \boldsymbol{v}}\Bigr|_{\substack{\boldsymbol{v}=\boldsymbol{v}^*\\t=t^*}}$ and $\frac{\partial \mathfrak{b}(\boldsymbol{v}^*,t^*)}{\partial t}:=\frac{\partial \mathfrak{b}( \boldsymbol{v},t)}{\partial t}\Bigr|_{\substack{\boldsymbol{v}=\boldsymbol{v}^*\\t=t^*}}$. For two sets $\mathcal{S}_1$ and $\mathcal{S}_2$, the Minkowski sum is defined as $\mathcal{S}_1\oplus \mathcal{S}_2:=\{\boldsymbol{v}_1+\boldsymbol{v}_2|\boldsymbol{v}_1\in \mathcal{S}_1, \boldsymbol{v}_2\in \mathcal{S}_2\}$. 

\subsection{Discontinuous Systems and Nonsmooth Analysis}
Consider $\dot{\boldsymbol{v}}=f(\boldsymbol{v},t)$ where $f:\mathbb{R}^d\times\mathbb{R}_{\ge 0}\to\mathbb{R}^d$ is locally bounded and measurable. We consider Filippov solutions \cite{filippov2013differential}  to this system and define the Filippov set-valued map 
\begin{align*}
F[f](\boldsymbol{v},t)&:=\overline{\text{co}}\{\lim_{i\to\infty}f(\boldsymbol{v}_i,t)|\boldsymbol{v}_i\to \boldsymbol{v},\boldsymbol{v}_i\notin N\cup N_f  \}
\end{align*}
where $\overline{\text{co}}$ denotes the convex closure; $N_f$ denotes the set of Lebesgue measure zero where $f(\boldsymbol{v},t)$ is discontinuous, while $N$ denotes an arbitrary set of Lebesgue measure zero. A Filippov solution to $\dot{\boldsymbol{v}}=f(\boldsymbol{v},t)$ is an absolutely continuous function $\boldsymbol{v}:[t_0,t_1]\to\mathbb{R}^d$ that satisfies $\dot{\boldsymbol{v}}(t)\in F[f](\boldsymbol{v},t)$ for almost all $t\in[t_0,t_1]$. Due to \cite[Prop.~3]{cortes2008discontinuous} it holds that there exists a Filippov solution to $\dot{\boldsymbol{v}}=f(\boldsymbol{v},t)$ if $f:\mathbb{R}^d\times\mathbb{R}_{\ge 0}\to\mathbb{R}^d$ is locally bounded and measurable. For switched systems with state-dependent switching, existence of Filippov solutions is discussed in \cite{leth2012formalism}. The switching mechanism of the switched system presented in Section~\ref{sec:strategy} is time-dependent so that \cite[Prop.~3]{cortes2008discontinuous} can still be applied. Consider a continuously differentiable function $\mathfrak{b}(\boldsymbol{v},t)$ so that Clarke's generalized gradient of $\mathfrak{b}(\boldsymbol{v},t)$ coincides with the gradient of $\mathfrak{b}(\boldsymbol{v},t)$ \cite[Prop.~6]{cortes2008discontinuous}, denoted by $\nabla \mathfrak{b}(\boldsymbol{v},t):=\begin{bmatrix}
{\frac{\partial \mathfrak{b}(\boldsymbol{v},t)}{\partial \boldsymbol{v}}} & {\frac{\partial \mathfrak{b}(\boldsymbol{v},t)}{\partial t}}
\end{bmatrix}$. The set-valued Lie derivative of $\mathfrak{b}(\boldsymbol{v},t)$ with respect to $F[f](\boldsymbol{v},t)$ at $(\boldsymbol{v},t)$ is then defined as 
\begin{align*}
\mathcal{L}_{F[f]}\mathfrak{b}(\boldsymbol{v},t)&:={\{\nabla \mathfrak{b}(\boldsymbol{v},t)}\begin{bmatrix}{\boldsymbol{\zeta}}^T & 1\end{bmatrix}^T|\boldsymbol{\zeta}\in F[f](\boldsymbol{v},t)\}.
\end{align*}
According to \cite[Thm.~2.2]{shevitz1994lyapunov}, it holds that 
\begin{align*}
\dot{\mathfrak{b}}(\boldsymbol{v}(t),t)\in\mathcal{L}_{F[f]}\mathfrak{b}(\boldsymbol{v}(t),t)
\end{align*} 
for almost all $t\in[t_0,t_1]$. Let $\hat{\mathcal{L}}_{F[f]}\mathfrak{b}(\boldsymbol{v},t):=\big\{\frac{\partial \mathfrak{b}(\boldsymbol{v},t)}{\partial \boldsymbol{v}}\boldsymbol{\zeta}|\boldsymbol{\zeta}\in F[f](\boldsymbol{v},t)\big\}$, the set-valued Lie derivative is then equivalent to $
\mathcal{L}_{F[f]}\mathfrak{b}(\boldsymbol{v},t)=\hat{\mathcal{L}}_{F[f]}\mathfrak{b}(\boldsymbol{v},t)\oplus \big\{\frac{\partial \mathfrak{b}(\boldsymbol{v},t)}{\partial t}\big\}$.
\begin{lemma}\label{lemma_liederivate}
Consider  $\dot{\boldsymbol{v}}=f_1(\boldsymbol{v},t)+f_2(\boldsymbol{v},t)$ where $f_1:\mathbb{R}^d\times\mathbb{R}_{\ge 0}\to\mathbb{R}^d$ and $f_2:\mathbb{R}^d\times\mathbb{R}_{\ge 0}\to\mathbb{R}^d$ are locally bounded and measurable. It then holds that 
\begin{align*}
\mathcal{L}_{F[f_1+f_2]}\mathfrak{b}(\boldsymbol{v},t)&\subseteq\\ &\hspace{-0.8cm}\hat{\mathcal{L}}_{F[f_1]}\mathfrak{b}(\boldsymbol{v},t)\oplus \hat{\mathcal{L}}_{F[f_2]}\mathfrak{b}(\boldsymbol{v},t)\oplus \Big\{\frac{\partial \mathfrak{b}(\boldsymbol{v},t)}{\partial t}\Big\}.
\end{align*}
\begin{IEEEproof}
Applying the definition of $\mathcal{L}_{F[f]}\mathfrak{b}(\boldsymbol{v},t)$ gives
\begin{align*}
\mathcal{L}_{F[f_1+f_2]}\mathfrak{b}(\boldsymbol{v},t)&\\
&\hspace{-2.55cm}:=\{{\nabla \mathfrak{b}(\boldsymbol{v},t)}\begin{bmatrix}{\boldsymbol{\zeta}}^T & 1\end{bmatrix}^T|\boldsymbol{\zeta}\in F[f_1+f_2](\boldsymbol{v},t)\} \\
&\hspace{-2.55cm}\subseteq {\{\nabla \mathfrak{b}(\boldsymbol{v},t)}\begin{bmatrix}{\boldsymbol{\zeta}}^T & 1\end{bmatrix}^T|\boldsymbol{\zeta}\in F[f_1](\boldsymbol{v},t)\oplus F[f_2](\boldsymbol{v},t)\}\\
&\hspace{-2.55cm} =\Big\{\frac{\partial \mathfrak{b}(\boldsymbol{v},t)}{\partial \boldsymbol{v}}\boldsymbol{\zeta}|\boldsymbol{\zeta}\in F[f_1](\boldsymbol{v},t)\oplus F[f_2](\boldsymbol{v},t)\Big\}\oplus\Big\{\frac{\partial \mathfrak{b}(\boldsymbol{v},t)}{\partial t}\Big\}\\
&\hspace{-2.55cm}=\hat{\mathcal{L}}_{F[f_1]}\mathfrak{b}(\boldsymbol{v},t) \oplus \hat{\mathcal{L}}_{F[f_2]}\mathfrak{b}(\boldsymbol{v},t)\oplus\Big\{\frac{\partial \mathfrak{b}(\boldsymbol{v},t)}{\partial t}\Big\}.
\end{align*}
where we used the fact that $F[f_1+f_2](\boldsymbol{v},t)\subseteq F[f_1](\boldsymbol{v},t)\oplus F[f_2](\boldsymbol{v},t)$ due to \cite[Thm. 1]{paden1987calculus}.
\end{IEEEproof}
\end{lemma}

\subsection{Signal Temporal Logic (STL)}
Signal temporal logic \cite{maler2004monitoring} is based on predicates $\mu$ that are obtained after evaluation of a continuously differentiable predicate function $h:\mathbb{R}^d\to\mathbb{R}$ as $\mu:=
\top$  if  $h(\boldsymbol{v})\ge 0$ and
 $\mu:=\bot$ if $h(\boldsymbol{v})< 0$ for $\boldsymbol{v}\in\mathbb{R}^d$. We consider, in this paper, an STL fragment that is recursively defined as
\begin{subequations}\label{eq:subclass}
\begin{align}
\psi \; &::= \; \top \; | \; \mu \; | \; \psi' \wedge \psi'' \label{eq:psi_class}\\
\phi \; &::= \;  G_{[a,b]}\psi \; | \; F_{[a,b]} \psi \;|\; \psi'  \until{a}{b} \psi'' \; | \; \phi' \wedge \phi''\label{eq:phi_class}
\end{align}
\end{subequations}
where $\psi'$, $\psi''$ denote formulas of class $\psi$ in \eqref{eq:psi_class}, whereas $\phi'$, $\phi''$ denote formulas of class $\phi$ in \eqref{eq:phi_class}. Note that $\neg \mu$ can be encoded in \eqref{eq:psi_class} by defining  $\bar{\mu}:=\neg \mu$ and  $\bar{h}(\boldsymbol{v}):=-h(\boldsymbol{v})$. The operators $\neg$, $\wedge$, $G_{[a,b]}$, $F_{[a,b]}$, and $\until{a}{b}$ denote the negation, conjunction, always, eventually, and until operators with $a\le b<\infty$. Formulas of class $\psi$ in \eqref{eq:psi_class} are non-temporal (Boolean)  formulas whereas formulas of class $\phi$ in \eqref{eq:phi_class} are temporal formulas. Let $(\boldsymbol{s},t)\models \phi$ denote the satisfaction relation, i.e., if a signal $\boldsymbol{s}:\mathbb{R}_{\ge 0}\to\mathbb{R}^d$ satisfies  $\phi$ at time $t$; $\phi$ is satisfiable if $\exists \boldsymbol{s}:\mathbb{R}_{\ge 0}\to\mathbb{R}^d$ such that $(\boldsymbol{s},0)\models \phi$. For a given $\boldsymbol{s}:\mathbb{R}_{\ge 0}\to\mathbb{R}^d$, the STL semantics \cite{maler2004monitoring} of the fragment in \eqref{eq:subclass} are recursively defined by: $(\boldsymbol{s},t) \models \mu$ iff $h(\boldsymbol{s}(t))\ge 0$, $(\boldsymbol{s},t) \models \psi' \wedge \psi''$ iff $(\boldsymbol{s},t) \models \psi' \wedge (\boldsymbol{s},t) \models \psi''$, $(\boldsymbol{s},t) \models G_{[a,b]}\psi$ iff $\forall \bar{t} \in[t+a,t+b] \text{, }(\boldsymbol{s},\bar{t})\models \psi$, $(\boldsymbol{s},t) \models F_{[a,b]}\psi$ iff $\exists \bar{t} \in[t+a,t+b] \text{ s.t. }(\boldsymbol{s},\bar{t})\models \psi$, and $(\boldsymbol{s},t) \models \psi' \until{a}{b} \psi''$ iff	$\exists \bar{t} \in[t+a,t+b]\text{ s.t. }(\boldsymbol{s},\bar{t})\models \psi''\wedge \forall \underline{t}\in[t,\bar{t}]\text{, }(\boldsymbol{s},\underline{t}) \models \psi'$.
%$(\boldsymbol{s},t) \models \phi' \wedge \phi''$ iff $(\boldsymbol{s},t) \models \phi' \wedge (\boldsymbol{s},t) \models \phi''$.
%\begin{align*}
%(\boldsymbol{s},t) \models \mu \;\; &\Leftrightarrow \;\; h(\boldsymbol{s}(t))\ge 0,\\
%(\boldsymbol{s},t) \models \neg\mu \;\; &\Leftrightarrow \;\; \neg((\boldsymbol{s},t) \models \mu),\\ 
%(\boldsymbol{s},t) \models \psi' \wedge \psi'' \;\; &\Leftrightarrow \;\; (\boldsymbol{s},t) \models \psi' \wedge (\boldsymbol{s},t) \models \psi'',\\
%(\boldsymbol{s},t) \models G_{[a,b]}\psi \;\; &\Leftrightarrow \;\; \forall \bar{t} \in[t+a,t+b] \text{,}(\boldsymbol{s},\bar{t})\models \psi, \\
%(\boldsymbol{s},t) \models F_{[a,b]}\psi \;\; &\Leftrightarrow \;\; \exists \bar{t} \in[t+a,t+b] \text{ s.t. }(\boldsymbol{s},\bar{t})\models \psi,\\
%(\boldsymbol{s},t) \models \psi' \until{a}{b} \psi''	 \;\; &\Leftrightarrow \;\; \exists \bar{t} \in[t+a,t+b]\text{ s.t. }(\boldsymbol{s},\bar{t})\models \psi'' \\
%& \hspace{0.6cm}\wedge \forall \underline{t}\in[t,\bar{t}]\text{,}(\boldsymbol{s},\underline{t}) \models \psi',\\
%(\boldsymbol{s},t) \models \phi' \wedge \phi'' \;\; &\Leftrightarrow \;\; (\boldsymbol{s},t) \models \phi' \wedge (\boldsymbol{s},t) \models \phi''.
%\end{align*} 
 Robust semantics \cite[Def. 3]{donze2} are recursively defined by
\begin{align*}
\rs{\mu}& := h(\boldsymbol{s}(t)), \;\;\; \rs{\neg\mu} := 	-\rs{\mu},\\
\rs{\psi' \wedge \psi''} &:= 	\min(\rs{\psi'},\rs{\psi''}),\\
%\rs{\phi_1 \vee \phi_2} &:= \max(\rs{\phi_1},\rs{\phi_2}),\\
\rs{G_{[a,b]} \psi} &:= \underset{\bar{t}\in[t+a,t+b]}{\min}\rss{\psi}{\bar{t}},\\
\rs{F_{[a,b]} \psi} &:= \underset{\bar{t}\in[t+a,t+b]}{\max}\rss{\psi}{\bar{t}},\\
\rs{\psi' \until{a}{b} \psi''} &:= \underset{\bar{t}\in [t+a,t+b]}{\max}  \min(\rss{\psi''}{\bar{t}}, \underset{\underline{t}\in[t,\bar{t}]}{\min}\rss{\psi'}{\underline{t}} ),  
\end{align*}
and determine how robustly a signal $\boldsymbol{s}$ satisfies $\phi$ at time $t$. It holds that $(\boldsymbol{s},t)\models \phi$ if $\rho^\phi(\boldsymbol{s},t)>0$ \cite[Prop.~16]{fainekos2009robustness}. 

\subsection{Control Barrier Functions encoding STL tasks}\label{sec:cbf_enc_stl}
Our previous work \cite{lindemann2018control} has established a connection between a function $\mathfrak{b}:\mathbb{R}^d\times \mathbb{R}_{\ge 0} \to \mathbb{R}$ (later shown to be a valid control barrier function) and the STL semantics of $\phi$ given in \eqref{eq:phi_class}. In particular, if this function is according to \cite[Steps A, B, and C]{lindemann2018control}, then, for a given signal $\boldsymbol{s}:\mathbb{R}_{\ge 0}\to \mathbb{R}^d$ with $\mathfrak{b}(\boldsymbol{s}(t),t)\ge 0$ for all $t\ge 0$, it holds that $(\boldsymbol{s},0)\models \phi$. Let
\begin{align*}
\mathfrak{C}(t):=\{\boldsymbol{v}\in \mathbb{R}^d | \mathfrak{b}(\boldsymbol{v},t)\ge 0\}
\end{align*}
so that equivalently $\boldsymbol{s}(t)\in\mathfrak{C}(t)$ for all $t\ge 0$ implies $(\boldsymbol{s},0)\models \phi$.  The conditions in \cite[Steps A, B, and C]{lindemann2018control} are summarized next. To encode conjunctions contained in $\phi$, a smooth approximation of the min-operator in the robust semantics is used. For $p$ functions $\mathfrak{b}_l:\mathbb{R}^d\times \mathbb{R}_{\ge 0} \to \mathbb{R}$ where $l\in\{1,\hdots,p\}$, let $\mathfrak{b}(\boldsymbol{v},t):=-\frac{1}{\eta}\ln\big(\sum_{l=1}^{p}\exp(-\eta \mathfrak{b}_l(\boldsymbol{v},t))\big)$ with $\eta>0$. Note that $\min_{l\in\{1,\hdots,p\}}\mathfrak{b}_l(\boldsymbol{v},t)\approx \mathfrak{b}(\boldsymbol{v},t)$ where the accuracy of this approximation increases with $\eta$, i.e.,
\begin{align*}
\lim_{\eta\to\infty}-\frac{1}{\eta}\ln\Big(\sum_{l=1}^{p} \exp(-\eta\mathfrak{b}_l(\boldsymbol{v},t))\Big)=\min_{l\in\{1,\hdots,p\}} \mathfrak{b}_l(\boldsymbol{v},t).
\end{align*}
Regardless of the choice of $\eta$, we have
\begin{align}\label{eq:under_approx}
-\frac{1}{\eta}\ln\Big(\sum_{l=1}^{p} \exp(-\eta\mathfrak{b}_l(\boldsymbol{v},t))\Big)\le \min_{l\in\{1,\hdots,p\}} \mathfrak{b}_l(\boldsymbol{v},t) %\\ 
%& \hspace{3cm}\le -\ln\Big(\sum_{i=1}^p \exp(-\mathfrak{b}_i(\boldsymbol{x},t))\Big) + \ln(p).
\end{align}
which is useful since $\mathfrak{b}(\boldsymbol{v},t)\ge 0$ implies  $\mathfrak{b}_l(\boldsymbol{v},t)\ge 0$ for each $l\in\{1,\hdots,p\}$, i.e., the conjunction operator can be encoded. Let the predicate funtion $h_l(\boldsymbol{v})$ correspond to the predicate $\mu_l$. In Steps A and B, we  illustrate the main idea for single temporal operators, i.e., only one always, eventually, or until operator is contained in $\phi$.
\begin{table}[h!]
\centering
\begin{tabular}{ |p{1.4cm}||p{6.2cm}|  }
 \hline
 \multicolumn{2}{|c|}{Step A) Single temporal operators in \eqref{eq:phi_class} \emph{without} conjunctions} \\
 \hline
 \hline
$G_{[a,b]}\mu_1$   & $p=1$, $\forall t^\prime\in[a,b]$, $\mathfrak{b}_1(\boldsymbol{v},t^\prime)\le h_1(\boldsymbol{v})$     \\
 \hline
$F_{[a,b]} \mu_1$ & $p=1$, $\exists t^\prime\in[a,b]$ s.t. $\mathfrak{b}_1(\boldsymbol{v},t^\prime)\le h_1(\boldsymbol{v})$  \\
 \hline
$\mu_1  \until{a}{b} \mu_2$ & $p=2$, $\exists t^\prime\in[a,b]$ s.t. $\mathfrak{b}_2(\boldsymbol{v},t^\prime)\le h_2(\boldsymbol{v})$, $\forall t^{\prime\prime}\in[0,t^\prime]$, $\mathfrak{b}_1(\boldsymbol{v},t^{\prime\prime})\le h_1(\boldsymbol{v})$\\
 \hline
\end{tabular} 
\end{table}

Note that \eqref{eq:under_approx} ensures satisfaction of $\mu_1  \until{a}{b} \mu_2$ if $\mathfrak{b}(\boldsymbol{s}(t),t)\ge 0$ for all $t\in[a,b]$. In Step B, we generalize the results from Step A, but now with conjunctions of predicates instead of a single predicate. Let $\psi_1:=\mu_1\wedge\hdots\wedge\mu_{\tilde{p}_1}$ and $\psi_2:=\mu_{\tilde{p}_1+1}\wedge\hdots\wedge\mu_{\tilde{p}_1+\tilde{p}_2}$ where $\tilde{p}_1,\tilde{p}_2\ge 1$.

\begin{table}[h!]
\centering
\begin{tabular}{ |p{1.4cm}||p{6.2cm}|  }
 \hline
 \multicolumn{2}{|c|}{Step B) Single temporal operators  in \eqref{eq:phi_class} \emph{with} conjunctions.} \\
 \hline
 \hline
$G_{[a,b]}\psi_1$  & $p=\tilde{p}_1$, $\forall t^\prime\in[a,b]$, $\forall l\in\{1,\hdots,\tilde{p}_1\}$, $\mathfrak{b}_l(\boldsymbol{v},t^\prime)\le h_l(\boldsymbol{v})$  \\
 \hline
$F_{[a,b]} \psi_1$ & $p=\tilde{p}_1$, $\exists t^\prime\in[a,b]$, $\forall l\in\{1,\hdots,\tilde{p}_1\}$  s.t. $\mathfrak{b}_l(\boldsymbol{v},t^\prime)\le h_l(\boldsymbol{v})$ \\
 \hline
$\psi_1  \until{a}{b} \psi_2$ & $p=\tilde{p}_1+\tilde{p}_2$, $\exists t^{\prime}\in[a,b]$, $\forall l^{\prime}\in\{\tilde{p}_1+1,\hdots,\tilde{p}_1+\tilde{p}_2\}$  s.t. $\mathfrak{b}_{l^{\prime}}(\boldsymbol{v},t^\prime)\le h_{l^{\prime}}(\boldsymbol{v})$ and $\forall t^{\prime\prime}\in[0,t^\prime]$, $\forall l^{\prime\prime}\in\{1,\hdots,\tilde{p}_1\}$, $\mathfrak{b}_{l^{\prime\prime}}(\boldsymbol{v},t^{\prime\prime})\le h_{l^{\prime\prime}}(\boldsymbol{v})$ \\
 \hline
\end{tabular} 
\end{table}

In Step C), conjunctions of single temporal operators are considered. The conditions on $\mathfrak{b}(\boldsymbol{v},t)$ are a straightforward extension of Steps A and B. For instance, consider $G_{[a_1,b_1]}\psi_1 \wedge F_{[a_2,b_2]} \psi_2 \wedge \psi_3  \until{a_3}{b_3} \psi_4$. Let $p=3$ where $\mathfrak{b}_1(\boldsymbol{v},t)$, $\mathfrak{b}_2(\boldsymbol{v},t)$, and $\mathfrak{b}_3(\boldsymbol{v},t)$ are associated with $G_{[a_1,b_1]}\psi_1$, $F_{[a_2,b_2]} \psi_2$, and $\psi_3  \until{a_3}{b_3} \psi_4$ and constructed as in Steps A and B.

Similar to \cite{lindemann2018control}, a switching mechanism is introduced and we integrate $\mathfrak{o}_l:\mathbb{R}_{\ge 0}\to\{0,1\}$ into $\mathfrak{b}(\boldsymbol{v},t):=-\frac{1}{\eta}\ln\big(\sum_{l=1}^p\mathfrak{o}_l(t)\exp(-\eta\mathfrak{b}_l(\boldsymbol{v},t))\big)$; $p$ is again the total number of functions $\mathfrak{b}_l(\boldsymbol{v},t)$ obtained from Steps A, B, and C and each $\mathfrak{b}_l(\boldsymbol{v},t)$ corresponds to either an always, eventually, or until operator with a corresponding time interval $[a_l,b_l]$. We remove single functions $\mathfrak{b}_l(\boldsymbol{v},t)$ from $\mathfrak{b}(\boldsymbol{v},t)$ when the corresponding always, eventually, or until operator is satisfied. For each temporal operator, the associated $\mathfrak{b}_l(\boldsymbol{v},t)$ is removed at $t=b_l$, i.e., $\mathfrak{o}_l(t)=1$ if $t<b_l$ and $\mathfrak{o}_l(t):=0$ if $t\ge b_l$. We denote the switching sequence by $\{s_0:=0,s_1,\hdots,s_q\}$ with $q\in\mathbb{N}$ as the total number of switches. This sequence is known due to knowledge of $[a_l,b_l]$. At time $t\ge s_j$ we have $s_{j+1}:= \text{argmin}_{b_l\in\{b_1,\hdots,b_p\}}\zeta(b_l,t)$ where $\zeta(b_l,t):=b_l-t$ if $b_l-t> 0$ and $\zeta(b_l,t):=\infty$ otherwise. We further require that each function $\mathfrak{b}_l(\boldsymbol{v},t)$ is continuously differentiable so that $\mathfrak{b}(\boldsymbol{v},t)$ is continuously differentiable on $\mathbb{R}^d\times (s_j,s_{j+1})$. 
\subsection{Coupled Multi-Agent Systems}

%A class $\mathcal{KL}$ function $\beta:\mathbb{R}_{\ge 0}\times\mathbb{R}_{\ge 0}\to\mathbb{R}_{\ge 0}$ is continuous, a class $\mathcal{K}$ function in its first argument, and decreasing in its second argument with $\beta(r,s)\to 0$ as $s\to\infty$ for each fixed $r$.
Consider $M$ agents modeled by an undirected graph $\mathcal{G}:=(\mathcal{V},\mathcal{E})$ where $\mathcal{V}:=\{1,\hdots,M\}$ while $\mathcal{E}\in \mathcal{V}\times \mathcal{V}$ indicates communication links. Let $\boldsymbol{x}_i\in\mathbb{R}^{n_i}$ and $\boldsymbol{u}_i\in \mathbb{R}^{m_i}$ be states and inputs of agent $i$. Also let $\boldsymbol{x}:=\begin{bmatrix} {\boldsymbol{x}_1}^T & \hdots & {\boldsymbol{x}_M}^T\end{bmatrix}^T\in\mathbb{R}^n$ with $n:=n_1+\hdots+n_M$. The dynamics of agent $i$ are 
\begin{align}\label{system_noise}
\dot{\boldsymbol{x}}_i&=f_i(\boldsymbol{x}_i,t)+g_i(\boldsymbol{x}_i,t)\boldsymbol{u}_i+c_i(\boldsymbol{x},t)
\end{align}
where $f_i:\mathbb{R}^{n_i}\times \mathbb{R}_{\ge 0}\to\mathbb{R}^{n_i}$, $g_i:\mathbb{R}^{n_i}\times \mathbb{R}_{\ge 0}\to\mathbb{R}^{n_i\times m_i}$, and $c_i:\mathbb{R}^{n}\times \mathbb{R}_{\ge 0}\to\mathbb{R}^{n_i}$ are locally Lipschitz continuous;  $c_i(\boldsymbol{x},t)$ may model \emph{given dynamical couplings} such as those induced by a mechanical connection between agents; $c_i(\boldsymbol{x},t)$ may also describe unmodeled dynamics or disturbances. We assume that $f_i(\boldsymbol{x}_i,t)$ and $g_i(\boldsymbol{x}_i,t)$ are only known by agent $i$ and $c_i(\boldsymbol{x},t)$ is bounded, but otherwise unknown so that no knowledge of $\boldsymbol{x}$ and $c_i(\boldsymbol{x},t)$ is required by agent $i$ for the control design. In other words, there exists a known $C\ge 0$ such that $\re{\|c_i(\boldsymbol{x},t)\|_\infty}\le C$ for all $(\boldsymbol{x},t)\in\mathbb{R}^{n}\times\mathbb{R}_{\ge 0}$. 
\begin{assumption}\label{ass1}
The function $g_i(\boldsymbol{x}_i,t)$ has full row rank for all $(\boldsymbol{x}_i,t)\in\mathbb{R}^{n_i}\times\mathbb{R}_{\ge 0}$.
\end{assumption}
\begin{remark}\label{rem:1}
Assumption \ref{ass1} implies $m_i\ge n_i$. Since $c_i(\boldsymbol{x},t)$ is not known by agent~$i$, the system \eqref{system_noise} is, however, \emph{not} feedback equivalent to $\dot{\boldsymbol{x}}_i=\boldsymbol{u}_i$. Canceling $f_i(\boldsymbol{x}_i)$ may also induce high control inputs, while we derive a minimum norm controller in Section \ref{sec:controller}.  Assumption~\ref{ass1} allows to decouple the construction of control barrier functions from the dynamics of the agents as discussed in Section \ref{sec:construction_rules}. In other words, for a function $\mathfrak{b}(\boldsymbol{x},t)$ it holds that ${\frac{\partial \mathfrak{b}(\boldsymbol{x},t)}{\partial \boldsymbol{x}_i}}g_i(\boldsymbol{x}_i,t)=\boldsymbol{0}$ if and only if ${\frac{\partial \mathfrak{b}(\boldsymbol{x},t)}{\partial \boldsymbol{x}_i}}=\boldsymbol{0}$. We note that most of the standard multi-agent literature assume simplified dynamics to deal with the complexity of the problem at hand. Collision avoidance, consensus, formation control, or connectivity maintenance can be achieved through a secondary controller $f_i^\text{u}$. Let  $\mathcal{V}_i^\text{u}\subseteq\mathcal{V}$ be a set of agents that \emph{induce dynamical couplings}, and let  $\boldsymbol{x}_i^\text{u}:=\begin{bmatrix} {\boldsymbol{x}_{j_1}}^T & \hdots & {\boldsymbol{x}_{j_{|\mathcal{V}_i^\text{u}|}}}^T\end{bmatrix}^T$ and $n_i^\text{u}:=n_{j_1}+\hdots+n_{j_{|\mathcal{V}_i^\text{u}|}}$ for $j_1,\hdots,j_{|\mathcal{V}_i^\text{u}|}\in\mathcal{V}_i^\text{u}$. By using $\boldsymbol{u}_i:={g_i(\boldsymbol{x}_i,t)}^T(g_i(\boldsymbol{x}_i,t){g_i(\boldsymbol{x}_i,t)}^T)^{-1}f_i^\text{u}(\boldsymbol{x}_i^\text{u},t)+\boldsymbol{v}_i$   the dynamics $\dot{\boldsymbol{x}}_i=f_i(\boldsymbol{x}_i,t)+f_i^\text{u}(\boldsymbol{x}^\text{u},t)+g_i(\boldsymbol{x}_i,t)\boldsymbol{v}_i+c_i(\boldsymbol{x},t)$ resemble \eqref{system_noise} if $f_i^\text{u}:\mathbb{R}^{n_i^\text{u}}\times\mathbb{R}_{\ge 0}\to\mathbb{R}^{n_i}$ is locally Lipschitz continuous.
\end{remark}

\subsection{Problem Formulation}

Consider $K$ temporal formulas $\phi_1,\hdots,\phi_K$ of the form \eqref{eq:phi_class} and let the satisfaction of $\phi_k$ for $k\in\{1,\hdots,K\}$ depend on the set of agents ${\mathcal{V}}_k\subseteq \mathcal{V}$. This means that knowledge of the solutions $\boldsymbol{x}_i:\mathbb{R}_{\ge 0}\to\mathbb{R}^{n_i}$ to \eqref{system_noise} for $i\in{\mathcal{V}}_k$ is sufficient to evaluate if $\phi_k$ is satisfied. Assume further that ${\mathcal{V}}_1\cup\hdots\cup{\mathcal{V}}_K=\mathcal{V}$ and that the sets of agents ${\mathcal{V}}_1,\hdots, {\mathcal{V}}_K\in\mathcal{V}$ are disjoint, i.e., ${\mathcal{V}}_{k_1}\cap{\mathcal{V}}_{k_2}=\emptyset$ for all $k_1,k_2\in\{1,\hdots,K\}$ with $k_1\neq k_2$. There are hence no formula dependencies between agents in $\mathcal{V}_{k_1}$ and agents in $\mathcal{V}_{k_2}$, although these agents may still be dynamically coupled through $c_i(\boldsymbol{x},t)$. The formula dependencies need  to be in accordance with the graph topology of $\mathcal{G}$ as follows.
\begin{assumption}\label{ass:com}
For each $\phi_k$ with $k\in\{1,\hdots,K\}$, it holds that $(i_1,i_2)\in\mathcal{E}$ for all $i_1,i_2\in{\mathcal{V}}_k$.
\end{assumption}

\begin{problem}\label{prob1}
Consider $K$ formulas $\phi_k$ of the form \eqref{eq:phi_class}. Derive a decentralized control law $\boldsymbol{u}_i$ for each agent $i\in\mathcal{V}$ so that, for each Filippov solution $\boldsymbol{x}:\mathbb{R}_{\ge 0}\to\mathbb{R}^{n}$ to \eqref{system_noise} under $\boldsymbol{u}_i$, $0<r\le \rho^{\phi_1 \wedge \hdots \wedge \phi_K}(\boldsymbol{x},0)$ where $r$ is maximized.
\end{problem}

\section{Barrier Function-based Control Strategies}
\label{sec:strategy}

We first motivate why the decentralized multi-agent case requires a discontinuous control law, while the centralized case, resembling a single-agent formulation, permits continuous control laws \cite[Coroll. 1]{lindemann2019control}.  For $j_1,\hdots,j_{|{\mathcal{V}}_k|}\in{\mathcal{V}}_k$, define $\bar{\boldsymbol{x}}_k:=\begin{bmatrix}
{\boldsymbol{x}_{j_1}}^T & \hdots & {\boldsymbol{x}_{j_{|{\mathcal{V}}_k|}}}^T
\end{bmatrix}^T\in \mathbb{R}^{\bar{n}_k}$ with $\bar{n}_k:=n_{j_1}+\hdots+n_{j_{|{\mathcal{V}}_k|}}$. Note that, for agents in $\mathcal{V}_k$, the stacked agent dynamics of the elements in \eqref{system_noise} are
\begin{align*}
\dot{\bar{\boldsymbol{x}}}_k=\bar{f}_k(\bar{\boldsymbol{x}}_k,t)+\bar{g}_k(\bar{\boldsymbol{x}}_k,t)\bar{u}_k(\bar{\boldsymbol{x}}_k,t)+\bar{c}_k(\boldsymbol{x},t)
\end{align*}
with $\bar{u}_k(\bar{\boldsymbol{x}}_k,t)$ explicitly depending on $\bar{\boldsymbol{x}}_k$ and $t$ and where
\begin{align*}
\bar{f}_k(\bar{\boldsymbol{x}}_k,t)&:=\begin{bmatrix}{f_{j_1}(\boldsymbol{x}_{j_1},t)}^T & \hdots & {f_{j_{|\mathcal{V}_k|}}(\boldsymbol{x}_{j_{|\mathcal{V}_k|}},t)}^T\end{bmatrix}^T, \\
\bar{g}_k(\bar{\boldsymbol{x}}_k,t)&:=\text{diag}\big({g_{j_1}(\boldsymbol{x}_{j_1},t)}, \hdots , {g_{j_{|\mathcal{V}_k|}}(\boldsymbol{x}_{j_{|\mathcal{V}_k|}},t)}\big),\\ 
\bar{\boldsymbol{u}}_k(\bar{\boldsymbol{x}}_k,t)&:=\begin{bmatrix}
{\boldsymbol{u}_{j_1}(\bar{\boldsymbol{x}}_k,t)}^T & \hdots & {\boldsymbol{u}_{j_{|{\mathcal{V}}_k|}}(\bar{\boldsymbol{x}}_k,t)}^T
\end{bmatrix}^T,\\
\bar{c}_k(\boldsymbol{x},t)&:=\begin{bmatrix}{c_{j_1}(\boldsymbol{x},t)}^T & \hdots & {c_{j_{|\mathcal{V}_k|}}(\boldsymbol{x},t)}^T\end{bmatrix}^T.
\end{align*} 
The function $\bar{c}_k(\boldsymbol{x},t)$ may dynamically couple some or even all agents. Let $\bar{\mathfrak{b}}_k:\mathbb{R}^{\bar{n}_k}\times\mathbb{R}_{\ge 0}\to\mathbb{R}$ denote the control barrier function corresponding to $\phi_k$ and accounting for Steps A, B, and C.  For the stacked agent dynamics, the centralized control barrier function condition (see, e.g., \cite[eq. (6)]{lindemann2019control}) is 
\begin{align}\label{eq:barrier_central}
\begin{split}
&\frac{\partial \bar{\mathfrak{b}}_k(\bar{\boldsymbol{x}}_k,t)}{\partial \bar{\boldsymbol{x}}_k}(\bar{f}_k(\bar{\boldsymbol{x}}_k,t)+\bar{g}_k(\bar{\boldsymbol{x}}_k,t)\bar{\boldsymbol{u}}_k(\bar{\boldsymbol{x}}_k,t)) \\
&\hspace{-0.5cm}+\frac{\partial \bar{\mathfrak{b}}_k(\bar{\boldsymbol{x}}_k,t)}{\partial t}\ge-\alpha_k(\bar{\mathfrak{b}}_k(\bar{\boldsymbol{x}}_k,t))+\re{\Big\|\frac{\partial \bar{\mathfrak{b}}_k(\bar{\boldsymbol{x}}_k,t)}{\partial \bar{\boldsymbol{x}}_k}\Big\|_1 C}
\end{split}
\end{align}
where $\alpha_k:\mathbb{R}\to\mathbb{R}$ is an extended class $\mathcal{K}$ function and where we use the constant  $C$  \re{since it holds that $\|\bar{c}_k(\boldsymbol{x},t)\|_\infty\le C$, which follows since $\|\bar{c}_k(\boldsymbol{x},t)\|_\infty= \max_{i\in \mathcal{V}_k} \|c_i(\boldsymbol{x},t)\|_\infty\le C$ since $\|c_i(\boldsymbol{x},t)\|_\infty\le C$ for each $i\in\mathcal{V}_k$ by assumption.} Note that  $\frac{\partial \bar{\mathfrak{b}}_k(\bar{\boldsymbol{x}}_k,t)}{\partial t}\ge -\alpha_k(\bar{\mathfrak{b}}_k(\bar{\boldsymbol{x}}_k,t))$ will hold if $\frac{\partial \bar{\mathfrak{b}}_k(\bar{\boldsymbol{x}}_k,t)}{\partial \bar{\boldsymbol{x}}_k}=\boldsymbol{0}$ as ensured by the control barrier function construction proposed in Section \ref{sec:construction_rules} and by virtue of Lemma~\ref{lemma:vCBF}; The solution of \eqref{eq:barrier_central}  admits a continuous and bounded control law $\bar{\boldsymbol{u}}_k(\bar{\boldsymbol{x}}_k,t)$ \cite[Coroll. 1]{lindemann2019control}. 
\begin{remark}\label{rem:pract_cent}
There are two ways to compute and implement $\bar{\boldsymbol{u}}_k(\bar{\boldsymbol{x}}_k,t)$ from \eqref{eq:barrier_central}: 1) Each agent $i\in\mathcal{V}_k$ solves \eqref{eq:barrier_central} and applies the portion $\boldsymbol{u}_i(\bar{\boldsymbol{x}}_k,t)$ of $\bar{\boldsymbol{u}}_k(\bar{\boldsymbol{x}}_k,t)$, or 2) Inequality \eqref{eq:barrier_central} is solved by one agent $i\in\mathcal{V}_k$ that sends the portions $\boldsymbol{u}_j(\bar{\boldsymbol{x}}_k,t)$ of $\bar{\boldsymbol{u}}_k(\bar{\boldsymbol{x}}_k,t)$ to the agents $j\in\mathcal{V}_k\setminus\{i\}$. The drawbacks  are that at least one agent needs to know the dynamics of each other agent, i.e., $\bar{f}_k(\bar{\boldsymbol{x}}_k,t)$ and $\bar{g}_k(\bar{\boldsymbol{x}}_k,t)$, and, for a large number of agents, \eqref{eq:barrier_central} may contain a large number of decision variables, equal to the dimension of $\bar{\boldsymbol{u}}_k(\bar{\boldsymbol{x}}_k,t)$. The second approach also requires more communication and lacks robustness since a malfunctioning agent $i$ results in a halt of the whole system. 
\end{remark}

We, as opposed to Remark \ref{rem:pract_cent}, propose the decentralization of \eqref{eq:barrier_central}, and hence of the control input computation, such that each agent computes its own control input $\boldsymbol{u}_i(\bar{\boldsymbol{x}}_k,t)$ alleviating the above issues. Each agent solves its own decentralized control barrier function condition so that their conjunction implies \eqref{eq:barrier_central}. A straightforward idea is to let each agent $i\in\mathcal{V}_k$ solve
\begin{align}\label{eq:barrier_decentral_wrong}
\begin{split}
&\frac{\partial \bar{\mathfrak{b}}_k(\bar{\boldsymbol{x}}_k,t)}{\partial \boldsymbol{x}_i}(f_i(\boldsymbol{x}_i,t)+g_i(\boldsymbol{x}_i,t)\boldsymbol{u}_i(\bar{\boldsymbol{x}}_k,t))\ge \\
&\hspace{0cm}-D_i\Big(\frac{\partial \bar{\mathfrak{b}}_k(\bar{\boldsymbol{x}}_k,t)}{\partial t}+\alpha_k(\bar{\mathfrak{b}}_k(\bar{\boldsymbol{x}}_k,t))\Big)+\re{\Big\|\frac{\partial \bar{\mathfrak{b}}_k(\bar{\boldsymbol{x}}_k,t)}{\partial \boldsymbol{x}_i}\Big\|_1  C}
\end{split}
\end{align}
where the weight $D_i:=\frac{1}{|\mathcal{V}_k|}$ distributes \eqref{eq:barrier_central} equally to each agent. Note that other weights $D_i$ could be imagined, as long as $\sum_{i\in\mathcal{V}_k} D_i=1$ similarly to \cite{wang2017safety}. We remark that we show, in the proof of Theorem \ref{theorem1}, why \eqref{eq:barrier_decentral_wrong} for each $i\in\mathcal{V}_k$ implies \eqref{eq:barrier_central}. With $D_i:=\frac{1}{|\mathcal{V}_k|}$, however, the obtained control law may induce problems when the gradients $\frac{\partial \bar{\mathfrak{b}}_k(\bar{\boldsymbol{x}}_k,t)}{\partial \boldsymbol{x}_i}$ become equal to the zero vector. In particular, assume that $\frac{\partial \bar{\mathfrak{b}}_k(\bar{\boldsymbol{x}}_k,t)}{\partial \boldsymbol{x}_i}= \boldsymbol{0}$ while $\exists j\in\mathcal{V}_k\setminus\{i\}$ such that $\frac{\partial \bar{\mathfrak{b}}_k(\bar{\boldsymbol{x}}_k,t)}{\partial \boldsymbol{x}_j}\neq \boldsymbol{0}$, then \eqref{eq:barrier_decentral_wrong} for agent $i$ may not be feasible and hence not imply \eqref{eq:barrier_central} (note in this case that $\frac{\partial \bar{\mathfrak{b}}_k(\bar{\boldsymbol{x}}_k,t)}{\partial \bar{\boldsymbol{x}}_k}\neq \boldsymbol{0}$). Severely critical, it can be seen that it may happen that $\|\boldsymbol{u}_i(\boldsymbol{x}_i,t)\|\to\infty$ as $\frac{\partial \bar{\mathfrak{b}}_k(\bar{\boldsymbol{x}}_k,t)}{\partial \boldsymbol{x}_i}\to \boldsymbol{0}$ if $\exists j\in\mathcal{V}_k\setminus\{i\}$ such that $\frac{\partial \bar{\mathfrak{b}}_k(\bar{\boldsymbol{x}}_k,t)}{\partial \boldsymbol{x}_j}\to \boldsymbol{v}$ for $\boldsymbol{v}\neq\boldsymbol{0}$. Consequently, a weight function $D_i:\mathbb{R}^{\bar{n}_k}\times\mathbb{R}_{\ge 0}$ is needed and, as it will turn out, this weight function will be discontinuous. 
\begin{remark}
Local Lipschitz continuity for barrier functions-based control laws has been proven in \cite[Thm. 3]{ames2017control} under the ``relative degree one condition". For \eqref{eq:barrier_decentral_wrong}, this condition is equivalent to $\frac{\partial \bar{\mathfrak{b}}_k(\bar{\boldsymbol{x}}_k,t)}{\partial \boldsymbol{x}_i}\neq \boldsymbol{0}$, which does not hold in general, so that discontinuities in the control law can be expected as analyzed in the proof of Theorem \ref{theorem1}. For \eqref{eq:barrier_central}, note that situations where $\frac{\partial \bar{\mathfrak{b}}_k(\bar{\boldsymbol{x}}_k,t)}{\partial \bar{\boldsymbol{x}}_k}= \boldsymbol{0}$ are taken into account in the proof of \cite[Coroll. 1]{lindemann2019control}, ensuring continuity of the control law.
\end{remark}

Section \ref{background_bf} extends \cite{lindemann2018control} and \cite[Coroll. 1]{lindemann2019control} to obtain a centralized control barrier function condition for multi-agent systems with discontinuous control laws. Section \ref{sec:controller} uses these results and proposes a control law, based on a decentralized control barrier function condition, that solves Problem \ref{prob1}. Sections \ref{background_bf} and \ref{sec:controller} assume the existence of the functions $\bar{\mathfrak{b}}_k(\bar{\boldsymbol{x}}_k,t)$ that satisfy Steps A, B, and C. In Section \ref{sec:construction_rules}, we present a procedure to construct such $\bar{\mathfrak{b}}_k(\bar{\boldsymbol{x}}_k,t)$.
\subsection{A Centralized Control Barrier Function Condition for Multi-Agent Systems with Discontinuous Control Laws}
\label{background_bf}

The results in this section are derived \emph{without} the need for Assumption \ref{ass1}. The functions $\bar{\mathfrak{b}}_k:\mathbb{R}^{\bar{n}_k}\times\mathbb{R}_{\ge 0}\to\mathbb{R}$ are continuously differentiable on $\mathbb{R}^{\bar{n}_k}\times(s_j^k,s_{j+1}^k)$ where $\{s_0^k:=0,s_1^k,\hdots,s_{q_k}^k\}$ are the associated switching sequences as discussed in Section \ref{sec:cbf_enc_stl}. Similarly, define 
\begin{align*}
\mathfrak{C}_k(t):=\big\{\bar{\boldsymbol{x}}_k\in \mathbb{R}^{\bar{n}_k} | \bar{\mathfrak{b}}_k(\bar{\boldsymbol{x}}_k,t)\ge 0\big\}.
\end{align*}   

For a particular $k\in\{1,\hdots,K\}$, let $\boldsymbol{x}:[t_j^k,t_{j+1}^k]\to\mathbb{R}^n$ be a Filippov solution to \eqref{system_noise} under the control laws $\boldsymbol{u}_i(\bar{\boldsymbol{x}}_k,t)$ where $t_j^k:=s_j^k$. We distinguish between $t_{j+1}^k$ and $s_{j+1}^k$ since we want to ensure closed-loop properties over $[s_j^k,s_{j+1}^k]$, while Filippov solutions may only be defined for $t_{j+1}^k<s_{j+1}^k$. 
\begin{definition}[Control Barrier Function]\label{cCBF}
The function $\bar{\mathfrak{b}}_k:\mathbb{R}^{\bar{n}_k}\times\mathbb{R}_{\ge 0}\to\mathbb{R}$ is a candidate control barrier function (cCBF) for $[s_j^k,s_{j+1}^k)$ if, for each $\bar{\boldsymbol{x}}_k(s_j^k)\in\mathfrak{C}_k(s_j^k)$, there exists an absolutely continuous function $\bar{\boldsymbol{x}}_k:[s_j^k,s_{j+1}^k)\to\mathbb{R}^{\bar{n}_k}$ such that $\bar{\boldsymbol{x}}_k(t)\in\mathfrak{C}_k(t)$ for all $t\in [s_j^k,s_{j+1}^k)$. A cCBF  $\bar{\mathfrak{b}}_k(\bar{\boldsymbol{x}}_k,t)$ for $[s_j^k,s_{j+1}^k)$ is a valid control barrier function (vCBF) for $[s_j^k,s_{j+1}^k)$ and for  \eqref{system_noise} under locally bounded and measurable control laws $\boldsymbol{u}_i(\bar{\boldsymbol{x}}_k,t)$ if the following holds. For each $i\in\mathcal{V}_k$ with $c_i:\mathbb{R}^n\times\mathbb{R}_{\ge 0}\to\mathbb{R}^{n_i}$ such that $\re{\|c_i(\boldsymbol{x},t)\|_\infty}\le C$, $\boldsymbol{x}(t_j^k)\in\mathbb{R}^n$ with $\bar{\boldsymbol{x}}_k(t_j^k)\in\mathfrak{C}_k(t_j^k)$ implies, for each Filippov solution $\boldsymbol{x}:[t_j^k,t_{j+1}^k]\to\mathbb{R}^n$ to \eqref{system_noise} under the control laws $\boldsymbol{u}_i(\bar{\boldsymbol{x}}_k,t)$ with $t_j^k=s_j^k$, that $\bar{\boldsymbol{x}}_k(t)\in\mathfrak{C}_k(t)$ for all $t\in[t_j^k,\min(t_{j+1}^k,s_{j+1}^k))$.
\end{definition}

Note that the definition of a vCBF does not require that $t_{j+1}^k\ge s_{j+1}^k$. In the remainder, we consider open sets $\mathfrak{D}_k\in\mathbb{R}^{\bar{n}_k}$ such that $\mathfrak{D}_k\supset\mathfrak{C}_k(t)$ for all $t\in[s_j^k,s_{j+1}^k)$. 

\begin{theorem}\label{theorem:disc_barrier}
Assume that $\bar{\mathfrak{b}}_k(\bar{\boldsymbol{x}}_k,t)$ is a cCBF for $[s_j^k,s_{j+1}^k)$. If each $\boldsymbol{u}_i(\bar{\boldsymbol{x}}_k,t)$ is locally bounded and measurable and if there exists an extended class $\mathcal{K}$ function $\alpha_k$ such that
\begin{align}\label{eq:barrier_ineq}
\begin{split}
&\min \mathcal{L}_{F[\bar{f}_k+\bar{g}_k\bar{\boldsymbol{u}}_k]}\bar{\mathfrak{b}}_k(\bar{\boldsymbol{x}}_k,t) \ge \\
&\hspace{2cm}-\alpha_k(\bar{\mathfrak{b}}_k(\bar{\boldsymbol{x}}_k,t))+\re{\Big\|\frac{\partial \bar{\mathfrak{b}}_k(\bar{\boldsymbol{x}}_k,t)}{\partial \bar{\boldsymbol{x}}_k}\Big\|_1 C}
\end{split}
\end{align} 
for all $(\bar{\boldsymbol{x}}_k,t)\in \mathfrak{D}_k\times (s_j^k,s_{j+1}^k)$, then $\bar{\mathfrak{b}}_k(\bar{\boldsymbol{x}}_k,t)$ is a vCBF for $[s_j^k,s_{j+1}^k)$ and for \eqref{system_noise} under $\boldsymbol{u}_i(\bar{\boldsymbol{x}}_k,t)$.

\begin{IEEEproof}
Note first that \eqref{eq:barrier_ineq} implies
\begin{align}\label{eq:theorem1_ineq}
\begin{split}
&\min \mathcal{L}_{F[\bar{f}_k+\bar{g}_k\bar{\boldsymbol{u}}_k]}\bar{\mathfrak{b}}_k(\bar{\boldsymbol{x}}_k,t) \oplus {\frac{\partial\bar{\mathfrak{b}}_k(\bar{\boldsymbol{x}}_k,t)}{\partial \bar{\boldsymbol{x}}_k}}\bar{c}_k(\boldsymbol{x},t)\ge \\
&\hspace{5.3cm}-\alpha_k(\bar{\mathfrak{b}}_k(\bar{\boldsymbol{x}}_k,t))
\end{split}
\end{align} 
since \re{$-\frac{\partial \bar{\mathfrak{b}}_k(\bar{\boldsymbol{x}}_k,t)}{\partial \bar{\boldsymbol{x}}_k}\bar{c}_k(\boldsymbol{x},t)\le \|{\frac{\partial \bar{\mathfrak{b}}_k(\bar{\boldsymbol{x}}_k,t)}{\partial \bar{\boldsymbol{x}}_k}}\|_1 \|\bar{c}_k(\boldsymbol{x},t)\|_\infty \le \|\frac{\partial \bar{\mathfrak{b}}_k(\bar{\boldsymbol{x}}_k,t)}{\partial \bar{\boldsymbol{x}}_k}\|_1 C$ by Hölder's inequality (recall also that $\|\bar{c}_k(\boldsymbol{x},t)\|_\infty \le C$ since $\|c_i(\boldsymbol{x},t)\|_\infty\le C$ for each $i\in\mathcal{V}_k$)}. Assume next that $\bar{\boldsymbol{x}}_k(t_j^k)\in\mathfrak{C}_k(t_j^k)$ and consider Filippov solutions $\boldsymbol{x}:[t_j^k,t_{j+1}^k]\to\mathbb{R}^n$ to \eqref{system_noise} under the control laws $\boldsymbol{u}_i(\bar{\boldsymbol{x}}_k,t)$ with $t_j^k=s_j^k$, which are ensured to exist since $f_i(\boldsymbol{x}_i,t)$, $g_i(\boldsymbol{x}_i,t)$, $c_i(\boldsymbol{x},t)$, and $\boldsymbol{u}_i(\bar{\boldsymbol{x}}_k,t)$ are locally bounded and measurable. Note hence that $\dot{\bar{\mathfrak{b}}}_k(\bar{\boldsymbol{x}}_k(t),t)\in\mathcal{L}_{F[\bar{f}_k+\bar{g}_k\bar{\boldsymbol{u}}_k+\bar{c}_k]}\bar{\mathfrak{b}}_k(\bar{\boldsymbol{x}}_k(t),t)$ for almost all $t\in(t_j^k,\min(t_{j+1}^k,s_{j+1}^k))$ and consequently also for almost all $t\in[t_j^k,\min(t_{j+1}^k,s_{j+1}^k))$. Due to \eqref{eq:theorem1_ineq} it holds that $\min \mathcal{L}_{F[\bar{f}_k+\bar{g}_k\bar{\boldsymbol{u}}_k]}\bar{\mathfrak{b}}_k(\bar{\boldsymbol{x}}_k(t),t) \oplus {\frac{\partial \bar{\mathfrak{b}}_k(\bar{\boldsymbol{x}}_k(t),t)}{\partial \bar{\boldsymbol{x}}_k}}\bar{c}_k(\boldsymbol{x}(t),t)\ge -\alpha_k(\bar{\mathfrak{b}}_k(\bar{\boldsymbol{x}}_k(t),t))$ and according to Lemma \ref{lemma_liederivate}, we have $\min \mathcal{L}_{F[\bar{f}_k+\bar{g}_k\bar{\boldsymbol{u}}_k+\bar{c}_k]}\bar{\mathfrak{b}}_k(\bar{\boldsymbol{x}}_k(t),t)\ge \min\{ \mathcal{L}_{F[\bar{f}_k+\bar{g}_k\bar{\boldsymbol{u}}_k]}\bar{\mathfrak{b}}_k(\bar{\boldsymbol{x}}_k(t),t) \oplus \hat{\mathcal{L}}_{F[\bar{c}_k]}\bar{\mathfrak{b}}_k(\bar{\boldsymbol{x}}_k(t),t)\}=\min \mathcal{L}_{F[\bar{f}_k+\bar{g}_k\bar{\boldsymbol{u}}_k]}\bar{\mathfrak{b}}_k(\bar{\boldsymbol{x}}_k(t),t)\oplus{\frac{\partial \bar{\mathfrak{b}}_k(\bar{\boldsymbol{x}}_k(t),t)}{\partial \bar{\boldsymbol{x}}_k}}\bar{c}_k(\boldsymbol{x}(t),t)$ since $\hat{\mathcal{L}}_{F[\bar{c}_k]}\bar{\mathfrak{b}}_k(\bar{\boldsymbol{x}}_k(t),t)=\{\frac{\partial \bar{\mathfrak{b}}_k(\bar{\boldsymbol{x}}_k(t),t)}{\partial \bar{\boldsymbol{x}}_k}\bar{c}_k(\boldsymbol{x}(t),t)\}$, i.e., a singleton, due to \cite[Thm.~1]{paden1987calculus} and since $\bar{\mathfrak{b}}_k(\bar{\boldsymbol{x}}_k,t)$ is continuously differentiable. It then holds that $\dot{\bar{\mathfrak{b}}}_k(\bar{\boldsymbol{x}}_k(t),t)\ge \min \mathcal{L}_{F[\bar{f}_k+\bar{g}_k\bar{\boldsymbol{u}}_k+\bar{c}_k]}\bar{\mathfrak{b}}_k(\bar{\boldsymbol{x}}_k(t),t) \ge -\alpha_k(\bar{\mathfrak{b}}_k(\bar{\boldsymbol{x}}_k(t),t))$. By \cite[Lem.~2]{glotfelter2017nonsmooth}, it follows that $\bar{\mathfrak{b}}_k(\bar{\boldsymbol{x}}_k(t),t)\ge 0$ for all $t\in[t_j^k,\min(t_{j+1}^k,s_{j+1}^k))$.
\end{IEEEproof}
\end{theorem}

\begin{remark}\label{rem:attrac}
We use $\mathfrak{D}_k$ to obtain a similar notion of robustness as discussed in \cite{xu2015robustness}. If $\bar{\boldsymbol{x}}_k(t)\in\mathfrak{D}_k\setminus\mathfrak{C}_k(t)$, note that $\dot{\bar{\mathfrak{b}}}_k(\bar{\boldsymbol{x}}_k(t),t)\ge -\alpha_k(\bar{\mathfrak{b}}_k(\bar{\boldsymbol{x}}_k(t),t))>0$ for almost all $t\in[s_j^k,s_{j+1}^k]$ since $\bar{\mathfrak{b}}_k(\bar{\boldsymbol{x}}_k(t),t)<0$, which is an important property in the experimental setup in Section \ref{sec:experiments}.
\end{remark}

To guarantee satisfaction of $\phi_k$, Filippov solutions $\boldsymbol{x}:[t_0:=0,t_1]\to\mathbb{R}^n$ need to be defined for $t_1\ge \max(s_q^1,\hdots,s_q^K)$ so that we require $\mathfrak{C}_k(t)$ to be compact. This requirement is not restrictive and can be achieved by considering $\phi_k\wedge \phi_k^\text{bd}$ instead of $\phi_k$, as assumed in the remainder, where $\phi_k^\text{bd}:=G_{[0,\infty)}\mu_k^\text{bd}$ with $h_k^\text{bd}(\bar{\boldsymbol{x}}_k):=D_k-\|\bar{\boldsymbol{x}}_k\|$ for a suitably selected $D_k\ge 0$.

\begin{corollary}\label{corollary_phi_sat}
Let $\bar{\mathfrak{b}}_k(\bar{\boldsymbol{x}}_k,t)$ satisfy the conditions in Steps A, B, and C for $\phi_k$ and be a cCBF for each  $[s_j^k,s_{j+1}^k)$. Let each $\boldsymbol{u}_i(\bar{\boldsymbol{x}}_k,t)$ be locally bounded and measurable. If, for each $k\in\{1,\hdots,K\}$, $\bar{\boldsymbol{u}}_k(\bar{\boldsymbol{x}}_k,t)$ is such that \eqref{eq:barrier_ineq} holds for all $(\bar{\boldsymbol{x}}_k,t)\in\mathfrak{D}_k\times(s_j^k,s_{j+1}^k)$, then it follows that $(\boldsymbol{x},0)\models \phi_1 \wedge \hdots\wedge \phi_K$ for each Filippov solution to \eqref{system_noise} under $\boldsymbol{u}_i(\bar{\boldsymbol{x}}_k,t)$.

\begin{IEEEproof}
Note that $\bar{\mathfrak{b}}_k(\bar{\boldsymbol{x}}_k,t)$ is piecewise continuous in $t$ with discontinuities at times $s_j^k$. The set $\mathfrak{C}_k(t)$ is non-decreasing at these switching times $s_j^k$, i.e., $\lim_{\tau\to_{-} s_j^{k}}\mathfrak{C}_k(\tau)\subseteq\mathfrak{C}_k(s_j^k)$ where $\lim_{\tau\to_{-} s_j^{k}}\mathfrak{C}_k(\tau)$ denotes the left-sided limit of  $\mathfrak{C}_k(t)$ at $t=s_j^k$. This follows due to the switching mechanism and, in particular, the function $\mathfrak{o}_l(t)$ as explained in Section \ref{sec:cbf_enc_stl}. It is hence sufficient to ensure forward invariance of $\mathfrak{C}_k(t)$ for each $[s_j^k,s_{j+1}^k)$ separately since $\bar{\boldsymbol{x}}_k(s_{j+1}^k)\in\mathfrak{C}_k(s_{j+1}^k)$ if  $\bar{\boldsymbol{x}}_k(t)\in\mathfrak{C}_k(t)$ for all $t \in[s_j^k,s_{j+1}^k)$. Due to Theorem~\ref{theorem:disc_barrier}, it follows that $\bar{\boldsymbol{x}}_k(t)\in\mathfrak{C}_k(t)$ for all $t\in[t_0,\min(t_1,s_1^1,\hdots,s_K^1))$. Note that  $\mathfrak{C}_k(t)\subseteq \{\bar{\boldsymbol{x}}_k\in\mathbb{R}^{\bar{n}_k}|D_k-\|\bar{\boldsymbol{x}}_k\|\ge 0\}$ and that  $\mathfrak{C}_k(t)\subset\mathfrak{D}_k$. Consequently, there exists a compact set $\mathfrak{D}_k'\subset\mathfrak{D}_k$ so that $\bar{\boldsymbol{x}}_k(t)\in\mathfrak{C}_k(t)$ implies $\bar{\boldsymbol{x}}_k(t)\in\mathfrak{D}_k'$. This means that $\boldsymbol{x}(t)$ remains in a compact set $\mathfrak{D}_1'\times\hdots\times\mathfrak{D}_K'$, which implies $t_1\ge \min(s_1^1,\hdots,s_K^1)$ by \cite[Ch.~2.7]{filippov2013differential}. The same reasoning can be applied for consecutive time intervals. By the conditions imposed on $\bar{\mathfrak{b}}_k(\bar{\boldsymbol{x}}_k,t)$ in Steps A, B, and C, it follows that each Filippov solution satisfies $(\bar{\boldsymbol{x}}_k,0)\models\phi_k$ since $\bar{\mathfrak{b}}_k(\bar{\boldsymbol{x}}_k(t),t)\ge 0$ for all $t\in[s_0^k,s_q^k]$ so that $(\boldsymbol{x},0)\models \phi_1 \wedge \hdots\wedge \phi_K$ follows.
\end{IEEEproof}
\end{corollary}

\subsection{Collaborative Control Laws based on a Decentralized Control Barrier Function Condition}
\label{sec:controller}

In this section, we again assume that Assumption \ref{ass1} holds. We first analyze cases where ${\frac{\partial \bar{\mathfrak{b}}_k(\bar{\boldsymbol{x}}_k,t)}{\partial \bar{\boldsymbol{x}}_k}}\bar{g}_k(\bar{\boldsymbol{x}}_k,t)= {\boldsymbol{0}}^T$. These cases mean that $\bar{\mathfrak{b}}_k(\bar{\boldsymbol{x}}_k,t)$, although possibly being a cCBF for $[s_j^k,s^k_{j+1})$, may not be a vCBF for $[s_j^k,s^k_{j+1})$ and for \eqref{system_noise} under any control law $\bar{\boldsymbol{u}}_k(\bar{\boldsymbol{x}}_k,t)$ since \eqref{eq:barrier_ineq} may fail to hold. Due to Assumption \ref{ass1}, it holds that the nullspace of  ${\bar{g}_k(\bar{\boldsymbol{x}}_k,t)}^T$ is empty, i.e., ${\frac{\partial \bar{\mathfrak{b}}_k(\bar{\boldsymbol{x}}_k,t)}{\partial \bar{\boldsymbol{x}}_k}}\bar{g}_k(\bar{\boldsymbol{x}}_k,t)= {\boldsymbol{0}}^T$ if and only if $\frac{\partial \bar{\mathfrak{b}}_k(\bar{\boldsymbol{x}}_k,t)}{\partial \bar{\boldsymbol{x}}_k}=\boldsymbol{0}$. To take care of these cases, we define 
\begin{align}
\omega_k(\bar{\boldsymbol{x}}_k,t)&:=\frac{\partial \bar{\mathfrak{b}}_k(\bar{\boldsymbol{x}}_k,t)}{\partial t}+\alpha_k(\bar{\mathfrak{b}}_k(\bar{\boldsymbol{x}}_k,t))\label{eq:omega}\\
\mathfrak{B}_{j}^k&:=\Big\{(\bar{\boldsymbol{x}}_k,t)\in\mathfrak{D}_k\times(s^k_j,s^k_{j+1})|\frac{\partial \bar{\mathfrak{b}}_k(\bar{\boldsymbol{x}}_k,t)}{\partial \bar{\boldsymbol{x}}_k}= \boldsymbol{0}\Big\}\nonumber
\end{align}
and pose the following assumption.

\begin{assumption}\label{ass3}
For some $\epsilon_k>0$, it holds that $\omega_k(\bar{\boldsymbol{x}}_k,t)\ge \epsilon_k$ for each $(\bar{\boldsymbol{x}}_k,t)\in\mathfrak{B}_{j}^k$. 
\end{assumption} 

 Assumption \ref{ass3} will be addressed in Section \ref{sec:construction_rules} in Lemma~\ref{lemma:vCBF} an the intuition  is that $\omega_k(\bar{\boldsymbol{x}}_k,t)\ge \epsilon_k>0$ ensures that \eqref{eq:barrier_ineq} can be satisfied by a proper choice of $\alpha_k$ even if $\frac{\partial \bar{\mathfrak{b}}_k(\bar{\boldsymbol{x}}_k,t)}{\partial \bar{\boldsymbol{x}}_k}= \boldsymbol{0}$. From now on, assume further that $\mathfrak{D}_k$ is bounded. 
\begin{theorem}\label{theorem1}
Let $\bar{\mathfrak{b}}_k(\bar{\boldsymbol{x}}_k,t)$ satisfy the conditions in Steps A, B, and C for $\phi_k$, be a cCBF for each $[s^k_j,s^k_{j+1})$, and satisfy Assumption \ref{ass3}. If, for each $k\in\{1,\hdots,K\}$, each agent $i\in{\mathcal{V}}_k$  applies the control law $\boldsymbol{u}_i(\bar{\boldsymbol{x}}_k,t):=\boldsymbol{u}_i$  where $\boldsymbol{u}_i$ is given by
\begin{subequations}\label{eq:qp_conv}
\begin{align}
&\underset{\boldsymbol{u}_i\in\mathbb{R}^{m_i}}{\operatorname{argmin}}\; \boldsymbol{u}_i^T\boldsymbol{u}_i\label{eq:const_qp_cost}\\
\begin{split}
\text{s.t. }&    \frac{\partial \bar{\mathfrak{b}}_k(\bar{\boldsymbol{x}}_k,t)}{\partial \boldsymbol{x}_i}(f_i(\boldsymbol{x}_i,t)+g_i(\boldsymbol{x}_i,t)\boldsymbol{u}_i)\ge\\
 & \hspace{0.8cm}-D_i(\bar{\boldsymbol{x}}_k,t)\omega_k(\bar{\boldsymbol{x}}_k,t)+\re{\Big\|{\frac{\partial \bar{\mathfrak{b}}_k(\bar{\boldsymbol{x}}_k,t)}{\partial \boldsymbol{x}_i}}\Big\|_1 }C,\label{eq:const_qp}
\end{split}
\end{align}
\end{subequations}
with\re{
\begin{align*}
D_i(\bar{\boldsymbol{x}}_k,t):=
\begin{cases}
\frac{\big\| \frac{\partial \bar{\mathfrak{b}}_k(\bar{\boldsymbol{x}}_k,t)}{\partial \boldsymbol{x}_i}\big\|_1}{\sum_{v\in\mathcal{V}_k}\big\| \frac{\partial \bar{\mathfrak{b}}_k(\bar{\boldsymbol{x}}_k,t)}{\partial \boldsymbol{x}_v}\big\|_1} &\hspace{-0.3cm}\text{if } \sum_{v\in\mathcal{V}_k}\big\| \frac{\partial \bar{\mathfrak{b}}_k(\bar{\boldsymbol{x}}_k,t)}{\partial \boldsymbol{x}_v}\big\|_1\neq 0\\
1 &\hspace{-0.3cm}\text{otherwise,}
\end{cases}
\end{align*} }
then it follows that $(\boldsymbol{x},0)\models \phi_1 \wedge \hdots\wedge \phi_K$ for each Filippov solution to \eqref{system_noise} under $\boldsymbol{u}_i(\bar{\boldsymbol{x}}_k,t)$. 

\begin{IEEEproof}
The proof can be found in the appendix.
\end{IEEEproof}
\end{theorem}

The load sharing function $D_i(\bar{\boldsymbol{x}}_k,t)$  shares the centralized control barrier function condition \eqref{eq:barrier_ineq} among agents by means of the decentralized control barrier function condition \eqref{eq:const_qp}. Computation of $\boldsymbol{u}_i$ is hence decentralized so that smaller optimization problems can be solved without the requirement that an agent knows $\bar{f}_k(\bar{\boldsymbol{x}}_k,t)$ and $\bar{g}_k(\bar{\boldsymbol{x}},t)$. The optimization program \eqref{eq:qp_conv} is a computationally tractable convex quadratic program with $m_i$ decision variables and agents need no knowledge of $\boldsymbol{x}$ and $c_i(\boldsymbol{x},t)$. Also, the desired robustness is obtained, e.g., even if an agent $i\in\mathcal{V}_k$ malfunctions, the other agents in $\mathcal{V}_k\setminus\{i\}$ will still work towards satisfying $\phi_k$.

\section{Control Barrier Function Construction}
\label{sec:construction_rules}

The construction of $\bar{\mathfrak{b}}_k(\bar{\boldsymbol{x}}_k,t)$ is the same for each $\phi_k$. For readability reasons, we hence omit the index $k$ and consider instead $\phi$ and $\mathfrak{b}(\boldsymbol{x},t)$ with $\boldsymbol{x}\in\mathbb{R}^n$.  To enforce the conditions in Steps A, B, and C, we will consider a function $\gamma_l:\mathbb{R}_{\ge 0}\to \mathbb{R}$ that is associated with the  predicate function $h_l(\boldsymbol{x})$ and the predicate $\mu_l$. Let $h^{\text{opt}}_l:=\sup_{\boldsymbol{x}\in\mathbb{R}^n} h_l(\boldsymbol{x})$ for which it has to hold that $h^{\text{opt}}_l\ge 0$. Otherwise, i.e, if $h^{\text{opt}}_l< 0$, $\mu_l$ is not satisfiable. We aim at satisfying $\phi$ with robustness $r\in\mathbb{R}_{\ge 0}$, i.e., $\rho^\phi(\boldsymbol{x},0)\ge r$, and proceed in two steps (Steps 1 and 2). Note that Steps A, B, and C lead to a function $\mathfrak{b}(\boldsymbol{x},t):=-\frac{1}{\eta}\ln\big(\sum_{l=1}^p\mathfrak{o}_l(t)\exp(-\eta\mathfrak{b}_l(\boldsymbol{x},t))\big)$ where each $\mathfrak{b}_l(\boldsymbol{x},t)$ is associated either with an eventually ($F_{[a_l,b_l]}\mu_l$) or an always ($G_{[a_l,b_l]}\mu_l$) formula. Recall that an until operator is encoded in Steps A, B, and C as the conjunction of an always and an eventually operator. We present in Step 1 how to construct $\mathfrak{b}(\boldsymbol{x},t)$ when $\phi:=F_{[a_l,b_l]}\mu_l$ or  $\phi:=G_{[a_l,b_l]}\mu_l$ where $\mu_l$ does not contain any conjunctions, i.e., $p=1$. In Step 2, we explain how to construct $\mathfrak{b}(\boldsymbol{x},t)$ in the more general case when $\phi$ contains conjunctions, i.e., $p>1$. 

\emph{Step 1)}  Consider $\phi:=G_{[a_l,b_l]}\mu_l$ or $\phi:=F_{[a_l,b_l]}\mu_l$ and let 
\begin{align}\label{eq:t_star}
t^*_l:=\begin{cases}
b_l &\text{if}\;\;\; F_{[a_l,b_l]} \mu_l\\
a_l &\text{if}\;\;\; G_{[a_l,b_l]} \mu_l,
\end{cases}
\end{align} 
which reflects the requirement that $\mu_l$ has to hold at least once between $[a_l,b_l]$ for $F_{[a_l,b_l]} \mu_l$ (here this time instant is chosen to be $t^*_l:=b_l$) or at all times within $[a_l,b_l]$ for $G_{[a_l,b_l]} \mu_l$ (indicated by $t^*_l:=a_l$). It is assumed that $b_l>0$. Otherwise, i.e., $b_l=0$, satisfaction of $\phi$ would purely depend on the initial condition of the system. Next, choose
\begin{align*}
r\in \begin{cases}
(0,h_l^\text{opt}) &\text{if } t_l^*>0\\
(0,h_l(\boldsymbol{x}(0))] & \text{if } t_l^*=0
\end{cases}
\end{align*}
where the second case is explained as follows: if $h_l(\boldsymbol{x}(0))< r$ and $t_l^*=0$, there does not exist a signal $\boldsymbol{x}:\mathbb{R}_{\ge 0}\to\mathbb{R}^n$ with an initial condition $\boldsymbol{x}(0)$  such that $\rho^\phi(\boldsymbol{x},0)\ge r$. Let now
\begin{align*}
\mathfrak{b}_l(\boldsymbol{x},t):=-\gamma_l(t)+h_l(\boldsymbol{x}).
\end{align*}
In \cite{lindemann2019decentralized}, $\gamma_l(t)$ is an exponential function. The drawback is, from a practical point of view, that larger control inputs may occur as compared to the case where $\gamma_l(t)$ is a linear function. We aim to avoid this and define the piecewise linear function
\begin{align*}
\gamma_l(t):=\begin{cases}
\frac{\gamma_{l,\infty}-\gamma_{l,0}}{t_l^*}t+\gamma_{l,0} &\text{if } t<t^*_l \\
\gamma_{l,\infty} &\text{otherwise. }\\
\end{cases}
\end{align*}
The switching sequence is now $\{s_0=0,s_1=b_l\}$ since $p=1$ and it holds that $\gamma_l(t)$ is continuous on $(s_0,s_1)$. We remark that $\gamma_l(t)$ is continuously differentiable on $(s_0,s_{1})$ if $\phi=F_{[a_l,b_l]}\mu_l$, while $\gamma_l(t)$ is only piecewise continuously differentiable on $(s_0,s_{1})$ if $\phi=G_{[a_l,b_l]}\mu_l$. In fact, in the latter case, $\gamma_l(t)$ is only continuously differentiable on $(s_0,t_l^*)$ and on $(t_l^*,s_1)$. This, however, does not affect the theoretical results derived in Corollary \ref{corollary_phi_sat} and Theorem \ref{theorem1}. To see this, note that $[s_0,s_{1})=[s_0,t^*_l)\cup[t^*_l,s_{1})$ and consider the modified switching sequence $\{\bar{s}_0:=s_0,\bar{s}_1=t_l^*,\bar{s}_2:=s_1\}$. Now, the same guarantees given in Corollary \ref{corollary_phi_sat} and Theorem~\ref{theorem1} apply for $\mathfrak{b}_l(\boldsymbol{x},t)$ under the modified switching sequence. Next, let 
\begin{subequations}\label{eq:gamma}
\begin{align}
\gamma_{l,0} &\in 
\big(-\infty,h_l(\boldsymbol{x}(0))\big)  \label{eq:gamma0}\\
\gamma_{l,\infty} &\in \big(\max(r,\gamma_{l,0}),h_l^\text{opt}\big)\label{eq:gammainfty}
\end{align}
\end{subequations}
so that $0\le\mathfrak{b}_l(\boldsymbol{x}(0),0)$ and $\mathfrak{b}_l(\boldsymbol{x}(0),0)\le h_l(\boldsymbol{x}(0))-r$ if $t^*_l=0$ so that a satisfaction with a robustness of $r$ is possible. By the choice of $\gamma_{l,\infty}$  it is ensured that $\mathfrak{b}_l(\boldsymbol{x}(t^\prime),t^\prime)\le h_l(\boldsymbol{x}(t^\prime))-r$ for all $t^\prime\ge t^*_l$. Hence, if now $\mathfrak{b}_l(\boldsymbol{x}(t^\prime),t^\prime)\ge 0$ for all $t^\prime\ge t^*_l$, then it follows that $ h_l(\boldsymbol{x}(t^\prime))-r\ge 0$, which implies $h_l(\boldsymbol{x}(t^\prime))\ge r$ leading to $\rho^{\phi}(\boldsymbol{x},0)\ge r$ by the choice of $t^*_l$ and $r$. We note that $\gamma_l(t)$ is a non-decreasing function. By these construction rules, it is straightforward to conclude that $\mathfrak{b}_l(\boldsymbol{x},t)$ is a cCBF for $[\bar{s}_0,\bar{s}_1)$ and for $[\bar{s}_1,\bar{s}_2)$. 

\begin{example}
Consider the formula $\phi:=G_{[7.5,10]}(\|\boldsymbol{x}\|<5)$ that yields $h_l(\boldsymbol{x}):=5-\|\boldsymbol{x}\|$ and $h_l^\text{opt}=5$ so that we can choose $r :=0.25$. Assume the initial condition $\boldsymbol{x}(0):=\begin{bmatrix}
5 & 5 \end{bmatrix}^T$ so that $h_l(\boldsymbol{x}(0))=-2.07$. We select   $t_l^*:=7.5$, $\gamma_{l,0}:=-2.5$, and $\gamma_{l,\infty}:=0.5$ in accordance with \eqref{eq:t_star} and \eqref{eq:gamma}. Recall that $\mathfrak{b}_l(\boldsymbol{x},t):=-\gamma_l(t)+h_l(\boldsymbol{x})$ and note that $\mathfrak{b}_l(\boldsymbol{x}(t),t)\ge 0$ for all $t\ge 0$ is equivalent to $h_l(\boldsymbol{x}(t))\ge \gamma_l(t)$ for all $t\ge 0$. This leads to $\rho^\phi(\boldsymbol{x},0)>r$, i.e., $(\boldsymbol{x},0)\models \phi$, by the construction of $\gamma_l(t)$ as illustrated in Fig. \ref{fig:principle_gamma}.

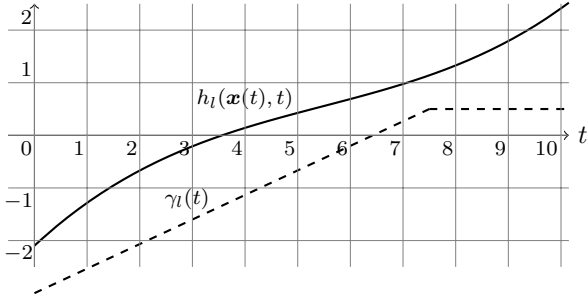
\begin{figure}
		\centering
			\begin{tikzpicture}[scale=0.7]
			\draw[->] (-0.5,0) -- (10.15,0) node[right] {$t$};
			\draw[->] (0,-2.5) -- (0,2.5) node[above] {};
			\draw[scale=1,domain=-0:7.5,smooth,variable=\x,dashed,thick] plot ({\x},{(0.5-(-3))/7.5*\x-3});
			\draw[scale=1,domain=7.5:10.15,smooth,variable=\x,dashed,thick] plot ({\x},{0.5});
			\draw[scale=1,domain=-0:10.15,smooth,variable=\x,thick] plot (\x,{-(-0.00736519*\x^3 + 0.121306*\x^2 - 0.926874*\x + 2.09632)});
			\node at (4,0.7) {\footnotesize $h_l(\boldsymbol{x}(t),t)$};
			\node at (2.9,-1.2) {\footnotesize $\gamma_l(t)$};
			\node at (-0.15,-0.25) {\footnotesize $0$};
			\node at (0.85,-0.25) {\footnotesize $1$};
			\node at (1.85,-0.25) {\footnotesize $2$};
			\node at (2.85,-0.25) {\footnotesize $3$};
			\node at (3.85,-0.25) {\footnotesize $4$};
			\node at (4.85,-0.25) {\footnotesize $5$};
			\node at (5.85,-0.25) {\footnotesize $6$};
			\node at (6.85,-0.25) {\footnotesize $7$};
			\node at (7.85,-0.25) {\footnotesize $8$};
			\node at (8.85,-0.25) {\footnotesize $9$};
			\node at (9.725,-0.25) {\footnotesize $10$};
			\node at (-0.15,1.25) {\footnotesize $1$};
			\node at (-0.3,-1.25) {\footnotesize $-1$};
			\node at (-0.15,2.25) {\footnotesize $2$};
			\node at (-0.3,-2.25) {\footnotesize $-2$};
			\draw[step=1,gray,very thin] (-0.5,-2.5) grid (10.15,2.5);
			\end{tikzpicture}\caption{The functions $\gamma_l(t)$ (dashed line) and $h_l(\boldsymbol{x}(t),t)$ (solid line) for $\phi:=G_{[7.5,10]}(\|\boldsymbol{x}\|<5)$ with $r:=0.25$ and a candidate trajectory $\boldsymbol{x}:\mathbb{R}_{\ge 0}\to\mathbb{R}^n$ satisfying $\phi$.}\label{fig:principle_gamma}
\end{figure}
\end{example}

\emph{Step 2)} For $p>1$, a more elaborate procedure is needed. Recall that $\mathfrak{b}(\boldsymbol{x},t):=-\frac{1}{\eta}\ln\big(\sum_{l=1}^p\mathfrak{o}_l(t)\exp(-\eta\mathfrak{b}_l(\boldsymbol{x},t))\big)$. Let, similarly to Step~1, $\mathfrak{b}_l(\boldsymbol{x},t):=-\gamma_l(t)+h_l(\boldsymbol{x})$ with $\gamma_l(t)$ according to \eqref{eq:gamma}. We further pose the following assumption.
\begin{assumption}\label{ass4}
Each predicate function contained in $\phi$, denoted by $h_l(\boldsymbol{x}):\mathbb{R}^n\to\mathbb{R}$ with $l\in\{1,\hdots,p\}$, is concave.
\end{assumption}

Concave predicate functions $h_l(\boldsymbol{x})$  contain the class of linear functions as well as functions that express, for instance, reachability tasks using predicates such as $\|\boldsymbol{x}-\boldsymbol{p}\|\le \epsilon$ for $\boldsymbol{p}\in\mathbb{R}^n$ and $\epsilon\in\mathbb{R}_{\ge 0}$. Assumption \ref{ass4} is needed to formally show that $\mathfrak{b}(\boldsymbol{x},t)$ is a cCBF and a vCBF (Lemmas \ref{lemma:cCBF} and \ref{lemma:vCBF}) relying on the fact that $\mathfrak{b}(\boldsymbol{x},t')$ is concave in $\boldsymbol{x}$ as proven next.
\begin{lemma}\label{lemma:concave}
Let Assumption~\ref{ass4} hold. Then, for a fixed $t'$, $\mathfrak{b}(\boldsymbol{x},t')$ is concave.

\begin{IEEEproof}
For a fixed $t'$,  $\eta\mathfrak{b}_l(\boldsymbol{x},t')$ is concave.  Due to \cite[Sec.~3.5]{boyd2004convex} it holds that $\exp(-\eta\mathfrak{b}_l(\boldsymbol{x},t'))$ is log-convex. It also holds that a sum of log-convex functions is log-convex. Hence, $-\frac{1}{\eta}\ln\big(\sum_{l=1}^p\mathfrak{o}_l(t)\exp(-\eta\mathfrak{b}_l(\boldsymbol{x},t'))\big)$ is concave.
\end{IEEEproof}
\end{lemma}

Compared to Step 1, it is now not enough to select $\gamma_{l,0}$ as in \eqref{eq:gamma0} to ensure $\mathfrak{b}(\boldsymbol{x}(0),0)\ge 0$ due to \eqref{eq:under_approx}.  To see this, consider $\mathfrak{b}(\boldsymbol{x},t):=-\frac{1}{\eta}\ln\big(\exp(-\eta\mathfrak{b}_1(\boldsymbol{x},t))+\exp(-\eta\mathfrak{b}_2(\boldsymbol{x},t))\big)$. If $\mathfrak{b}_1(\boldsymbol{x}(0),0)> 0$ and $\mathfrak{b}_2(\boldsymbol{x}(0),0)> 0$ (which is both ensured by \eqref{eq:gamma0}), then it does not neccessarily hold that $\mathfrak{b}(\boldsymbol{x}(0),0)\ge 0$ depending on the value of $\eta$. Therefore, $\eta$ now needs to be selected sufficiently large. Note again that increasing $\eta$ increases the accuracy of the approximation used for conjunctions. More importantly,  $\gamma_{l,\infty}$, which has to be selected according to  \eqref{eq:gammainfty}, and $r$ need to be selected so that for all $t\in[s_0,s_{q}]$ there exists $\boldsymbol{x}\in\mathbb{R}^n$ so that $\mathfrak{b}(\boldsymbol{x},t)\ge 0$. Define next 
$\boldsymbol{\gamma}_{0}:=\begin{bmatrix} \gamma_{1,0} & \hdots & \gamma_{p,0}\end{bmatrix}^T$ and $\boldsymbol{\gamma}_{\infty}:=\begin{bmatrix} \gamma_{1,\infty} & \hdots & \gamma_{p,\infty}\end{bmatrix}^T$  that contain the parameters $\gamma_{l,0}$ and $\gamma_{l,\infty}$ for each eventually- and always-operator encoded in $\mathfrak{b}_l(\boldsymbol{x},t)$. Let $\boldsymbol{\xi}_1,\hdots,\boldsymbol{\xi}_q\in \mathbb{R}^n$ and define $\boldsymbol{\xi}:=\begin{bmatrix}
{\boldsymbol{\xi}_1}^T & \hdots & {\boldsymbol{\xi}_q}^T
\end{bmatrix}^T$. As argued in Section \ref{background_bf}, $\mathfrak{C}(t):=\{\boldsymbol{x}\in \mathbb{R}^{n} | \mathfrak{b}(\boldsymbol{x},t)\ge 0\}$ needs to be compact. This is realized by including $\mathfrak{b}_{p+1}(\boldsymbol{x},t):=D-\|\boldsymbol{x}\|$ and $\mathfrak{o}_{p+1}(t):=1$ into $\mathfrak{b}(\boldsymbol{x},t):=-\frac{1}{\eta}\ln\big(\sum_{l=1}^{p+1}\mathfrak{o}_l(t)\exp(-\eta\mathfrak{b}_l(\boldsymbol{x},t))\big)$ for a suitably selected $D$. Select  $\eta$, $r$, $D$, $\boldsymbol{\gamma}_{0}$, and $\boldsymbol{\gamma}_{\infty}$ according to the solution of the following optimization problem
\begin{subequations}\label{eq:cbf_selection}
\begin{align}
&\underset{\eta,r,D,\boldsymbol{\gamma}_{0},\boldsymbol{\gamma}_{\infty},\boldsymbol{\xi}}{\operatorname{argmax}} r\label{optim_a}\\
\text{s.t.} \; &\mathfrak{b}(\boldsymbol{x}(0),0)\ge \chi \label{optim_b}\\
&\lim_{\tau\to s_j^-}\mathfrak{b}(\boldsymbol{\xi}_j,\tau)\ge \chi  \;\;\; \text{for each } j\in\{1,\hdots,q \} \label{optim_c}\\
& \gamma_{l,0} \text{ as in } \eqref{eq:gamma0} \text{ for each } l\in\{1,\hdots,p\}\label{optim_d}\\
& \gamma_{l,\infty} \text{ as in } \eqref{eq:gammainfty} \text{ for each } l\in\{1,\hdots,p\}\label{optim_e}\\
& \eta> 0 \;\; \text{and} \;\; r> 0 \;\; \text{and} \;\; D>0\label{optim_f}.
\end{align}
\end{subequations}
where $\chi\ge 0$ is a given parameter. Note that $\lim_{\tau\to s_j^-}\mathfrak{b}(\boldsymbol{\xi}_j,\tau)$ can easily be evaluated since $\mathfrak{o}_l(t)$ is piecewise continuous.  
\begin{remark}
The optimization problem \eqref{eq:cbf_selection} is nonconvex. An MILP formulation such as in \cite{raman1} provides, for discrete-time systems, an open-loop control sequence that needs to be iteratively solved  online in order to get a feedback control law. We obtain in \eqref{eq:cbf_selection}, which can be solved offline, a control barrier function that can be used, in a provably correct manner, to obtain a continuous feedback control law as in \eqref{eq:qp_conv}. Compared to \cite{lindemann2019decentralized}, we observed faster computation times due to the use of piecewise linear functions $\gamma_l(t)$ instead of exponential ones. If maximization of $r$ is not of interest, then a feasibility program with the constraints in \eqref{optim_b}-\eqref{optim_f} can be solved instead. 
\end{remark}

Denote the modified switching sequence by $\{\bar{s}_0:=0,\bar{s}_1,\hdots,\bar{s}_{\bar{q}}=s_q\}$ where $\bar{q}$ denotes the number of switches  with $\bar{q}\ge q$. More formally, let $\boldsymbol{t}:=\{
a_{l_1}, \hdots, a_{l_{\bar{p}}},b_1, \hdots, b_p
\}$ where, for $a_l\in\{a_{l_1}, \hdots, a_{l_{\bar{p}}}\}$, the corresponding $\mathfrak{b}_l(\boldsymbol{x},t)$ encodes an always operator, i.e., $\bar{p}$ denotes the number of $\mathfrak{b}_l(\boldsymbol{x},t)$ in $\mathfrak{b}(\boldsymbol{x},t)$ that encode an always operator. At time $t\ge\bar{s}_j$, we  define $\bar{s}_{j+1}:=\text{argmin}_{t^*\in\boldsymbol{t}}\zeta(t^*,t)$ with $\zeta(t^*,t):=t^*-t$ if $t^*-t> 0$ and $\zeta(t^*,t):=\infty$ otherwise. We now show that  $\mathfrak{b}(\boldsymbol{x},t)$ is a cCBF for each  $[\bar{s}_{j},\bar{s}_{j+1})$. 
\begin{lemma}\label{lemma:cCBF}
Let Assumption \ref{ass4} hold. Then the function $\mathfrak{b}(\boldsymbol{x},t)$ obtained by the solution of \eqref{eq:cbf_selection} is a cCBF for each $[\bar{s}_{j},\bar{s}_{j+1})$.

\begin{IEEEproof}
Feasibility of \eqref{eq:cbf_selection} implies that $\mathfrak{C}(t)$ is non-empty for all $t\in[s_0,s_q]$. This follows due to \eqref{optim_b}, \eqref{optim_c}, and since $\mathfrak{b}(\boldsymbol{x},t)$ is non-increasing in $t$ for all $t\in[s_j, s_{j+1})$ by \eqref{optim_d}-\eqref{optim_e}, which implies $\mathfrak{C}(t_1)\supseteq\mathfrak{C}(t_2)$ for $s_j\le t_1<t_2<s_{j+1}$. For $\mathfrak{b}(\boldsymbol{x},t)$ to be a cCBF for $[\bar{s}_j,\bar{s}_{j+1})$, there needs to exist an absolutely continuous function $\boldsymbol{x}:[\bar{s}_j,\bar{s}_{j+1})\to\mathbb{R}^n$ for each $\boldsymbol{x}(s_j)\in\mathfrak{C}(s_j)$ such that $\boldsymbol{x}(t)\in\mathfrak{C}(t)$ for all $t\in[\bar{s}_j,\bar{s}_{j+1})$. Since $\mathfrak{b}(\boldsymbol{x},t)$ is concave in $\boldsymbol{x}$ for each fixed $t$, it holds that all superlevel sets of $\mathfrak{b}(\boldsymbol{x},t)$ are convex \cite[Sec. 3.1.6]{boyd2004convex} and hence $\mathfrak{C}(t)$ is connected. Since $\frac{\partial \mathfrak{b}(\boldsymbol{x},t)}{\partial t}$ is finite, the existence of an absolutely continuous function $\boldsymbol{x}:[\bar{s}_j,\bar{s}_{j+1})\to\mathbb{R}^n$ such that $\mathfrak{b}(\boldsymbol{x}(t),t)\ge 0$ for all $t\in[\bar{s}_j,\bar{s}_{j+1})$ follows.
\end{IEEEproof}
\end{lemma}

Lemma \ref{lemma:cCBF} has shown that $\mathfrak{b}(\boldsymbol{x},t)$ is a cCBF, while we next show that $\alpha$ can be selected such that $\mathfrak{b}(\boldsymbol{x},t)$ is a vCBF.
% and define $\mathfrak{C}_{\chi}(t):=\{\boldsymbol{x}\in\mathbb{R}^n|\mathfrak{b}(\boldsymbol{x},t)\ge \chi\}$.
\begin{lemma}\label{lemma:vCBF}
Assume that \eqref{eq:cbf_selection} is solved for $\chi>0$, then $\alpha$ can be selected such that $\mathfrak{b}(\boldsymbol{x},t)$ satisfies Assumption \ref{ass3}.

\begin{IEEEproof}
Concavity of $\mathfrak{b}(\boldsymbol{x},t)$ in $\boldsymbol{x}$ implies that, for each $t'\in[s_j,s_{j+1})$,  $\boldsymbol{x}^{*}_{t'}:=\text{argmax}_{\boldsymbol{x}\in\mathbb{R}^n}\mathfrak{b}(\boldsymbol{x},t')$ is such that $\boldsymbol{x}^{*}_{t'}\in\mathfrak{C}(t')$ (recall that $\chi>0$) with $\mathfrak{b}(\boldsymbol{x}^*_{t'},t')>\mathfrak{b}(\boldsymbol{x},t')$ for all $\boldsymbol{x}\neq\boldsymbol{x}^*_{t'}$. Furthermore,  $\frac{\partial \mathfrak{b}(\boldsymbol{x}',t')}{\partial \boldsymbol{x}}=\boldsymbol{0}$ if and only if $\boldsymbol{x}':=\boldsymbol{x}^{*}_{t'}$. It holds that $\mathfrak{b}(\boldsymbol{x}^{*}_{t'},t')\ge \chi>0$ for each $t'\in[s_0,s_{q}]$ due to \eqref{optim_b} and \eqref{optim_c} so that $\mathfrak{b}_l(\boldsymbol{x}^{*}_{t'},t')\ge \chi>0$ for each $l\in\{1,\hdots,p+1\}$ with $\mathfrak{o}_l(t')=1$. Next, note that there exists a constant $\mathfrak{b}_l^\text{max}$  for each $l\in\{1,\hdots,p+1\}$ such that $\mathfrak{b}_l(\boldsymbol{x}^*_{t'},t')\le \mathfrak{b}_l^\text{max}$ for each $t'\in[s_0,s_q]$  due to continuity of $\mathfrak{b}_l(\boldsymbol{x},t)$ on $\mathfrak{D}\times[s_0,s_q]$. Let $\mathfrak{b}^\text{max}:=\max(\mathfrak{b}_1^\text{max},\hdots,\mathfrak{b}_{p+1}^\text{max})$ so that  $\max(\mathfrak{b}_1(\boldsymbol{x}^{*}_{t'},t'),\hdots,\mathfrak{b}_{p+1}(\boldsymbol{x}^{*}_{t'},t'))\le \mathfrak{b}^\text{max}$ and let $\Delta_l:=\sup_{t\ge 0}|\frac{\partial \mathfrak{b}_l(\boldsymbol{x},t)}{\partial t}|=\frac{\gamma_{l,\infty}-\gamma_{l,0}}{t_l^*}$. Hence, it follows that
\begin{align*}
\frac{\partial \mathfrak{b}(\boldsymbol{x}^*_{t'},t')}{\partial t}&=\frac{\sum_{l=1}^{p+1}\mathfrak{o}_l(t')\exp(-\eta\mathfrak{b}_l(\boldsymbol{x}^*_{t'},t'))\frac{\partial \mathfrak{b}_l(\boldsymbol{x}^*_{t'},t')}{\partial t}}{\sum_{l=1}^{p+1}\mathfrak{o}_l(t')\exp(-\eta\mathfrak{b}_l(\boldsymbol{x}^*_{t'},t'))}\\
&=\frac{-\sum_{l=1}^{p+1}\exp(-\eta\mathfrak{b}_l(\boldsymbol{x}^*_{t'},t'))|\frac{\partial \mathfrak{b}_l(\boldsymbol{x}^*_{t'},t')}{\partial t}|}{\sum_{l=1}^{p+1}\exp(-\eta\mathfrak{b}_l(\boldsymbol{x}^*_{t'},t'))}\\
&\ge \frac{-\exp(-\eta\chi)\Delta_l}{\exp(-\eta\mathfrak{b}^\text{max})}=:\zeta.
\end{align*}
where $\zeta$ is negative. If it is now guaranteed that $\zeta\ge -\alpha(\chi)+\epsilon$,  it holds that $\frac{\partial \mathfrak{b}(\boldsymbol{x}^*_{t'},t')}{\partial t}\ge -\alpha(\mathfrak{b}(\boldsymbol{x}^*_{t'},t')+\epsilon$ for all $t'\in[s_0,s_q]$ so that Assumption \ref{ass3} holds. By the specific choice of $\alpha(\chi)=\kappa\chi$, we can select $\kappa\ge \frac{\epsilon-\zeta}{\chi}$ such that this is the case.
\end{IEEEproof}
\end{lemma}

The intuition behind Lemma \ref{lemma:vCBF} is that $\chi>0$ ensures that $\mathfrak{b}(\boldsymbol{x},t)\ge\chi$ if $\frac{\partial \mathfrak{b}(\boldsymbol{x},t)}{\partial \boldsymbol{x}}=\boldsymbol{0}$ and that then choosing $\kappa$ in $\alpha(\chi)=\kappa\chi$ large enough guarantees that Assumption \ref{ass3} holds. We combine the results from Sections \ref{sec:controller} and \ref{sec:construction_rules}.
\begin{theorem}\label{corr3}
Consider the same assumptions as in Theorem~\ref{theorem1}. If each $\phi_k$ additionally satisfies Assumption  \ref{ass4}, $\bar{\mathfrak{b}}_k(\bar{\boldsymbol{x}}_k,t)$ is the solution of \eqref{eq:cbf_selection} for $\chi>0$, and $\alpha_k(\chi):=\kappa\chi$ with $\kappa>\frac{\epsilon-\zeta}{\chi}$, then  $\rho^{\phi_k}(\bar{\boldsymbol{x}}_k,0)\ge r_k> 0$ where $r_k$ is obtained by the solution of \eqref{eq:cbf_selection} for each $k\in\{1,\hdots,K\}$. 

\begin{IEEEproof}
Follows by Theorem \ref{theorem1} and Lemmas \ref{lemma:cCBF} and \ref{lemma:vCBF}.
\end{IEEEproof}
\end{theorem}

\section{Experiments}
\label{sec:experiments}

%\begin{figure}
%\centering
%\includegraphics[scale=0.055]{figures/robots}
%\caption{Nexus 4WD Mecanum Robotic Cars. The yellow base of each robot is equipped with a black robotic arm on top. }
%\label{fig:robots}
%\end{figure}

We consider three Nexus 4WD Mecanum Robotic Cars, which are equipped with low-level PID controllers that track translational and rotational velocity commands. The state of robot $i$ is $\boldsymbol{x}_i:=\begin{bmatrix}
x_i & y_i & \theta_i
\end{bmatrix}^T$ where $\boldsymbol{p}_i:=\begin{bmatrix} x_i & y_i\end{bmatrix}^T$ denotes the two dimensional position while $\theta_i$ denotes the orientation. For simplicity, we here assume that all states are given in a global coordinate frame. Conversion from local to global coordinate frames is performed by each robot where the local information is obtained by means of a motion capture system. The considered dynamics are given by $\dot{\boldsymbol{x}}_i=\boldsymbol{u}_i+f_i^\text{u}(\boldsymbol{x},t)+c_i(\boldsymbol{x},t)$
where $f_i^\text{u}(\boldsymbol{x},t)$ describes induced dynamical couplings as discussed in Remark \ref{rem:1}, here used for the purpose of collision avoidance. In particular, $f_i^\text{u}(\boldsymbol{x},t)$ is a potential field inducing a repulsive force between two robots when the distance between them is below $0.65$ meters; $c_i(\boldsymbol{x},t)$ models disturbances such as those induced by the digital implementation of the continuous-time control law or inaccuracies in the low-level PID controllers with $C:=2$. The robots are subject to  $\phi:=\phi'\wedge\phi''\wedge\phi'''\wedge\phi''''$ with
 \begin{align*}
 \phi'&:=G_{[15,90]}(\|\boldsymbol{p}_1+\boldsymbol{p}_x-\boldsymbol{p}_2\|\le \epsilon)\\
 \phi''&:=G_{[25,35]}(\|\boldsymbol{p}_1+\boldsymbol{p}_y-\boldsymbol{p}_3\|\le \epsilon)\\
 &\hspace{4cm}\wedge F_{[30,35]}\,(\|\boldsymbol{p}_1-\boldsymbol{p}_B\|\le \epsilon)\\
  \phi'''&:=F_{[40,60]}(\|\boldsymbol{p}_3-\boldsymbol{p}_C\|\le \epsilon)\\
 \phi''''&:=F_{[50,90]}\big((\|\boldsymbol{p}_1-\boldsymbol{p}_A\|\le \epsilon)\wedge(\|\boldsymbol{p}_2+\boldsymbol{p}_x-\boldsymbol{p}_3\|\le \epsilon)\big)
 \end{align*}
where $\epsilon:=0.33$, $\boldsymbol{p}_A:=\begin{bmatrix} -1.2 & 1.2 \end{bmatrix}^T$, $\boldsymbol{p}_B:=\begin{bmatrix} 1.2 & 1.2 \end{bmatrix}^T$, $\boldsymbol{p}_C:=\begin{bmatrix} 1.2 & -1.2 \end{bmatrix}^T$, $\boldsymbol{p}_x:=\begin{bmatrix} 0.8 & 0 \end{bmatrix}^T$, $\boldsymbol{p}_y:=\begin{bmatrix} 0 & -0.8 \end{bmatrix}^T$.

The software implementation is available under \cite{code_repository} (also including a detailed description of $f_i^\text{u}(\boldsymbol{x},t)$), written in \emph{C++}, and embedded in the \emph{Robot Operating System} (ROS) \cite{quigley2009ros}. The quadratic program \eqref{eq:qp_conv} is solved using \emph{CVXGEN} \cite{mattingley2012cvxgen} at a frequency of $50$ Hz;  $\mathfrak{b}(\boldsymbol{x},t)$ corresponding to $\phi$ is obtained offline and in \emph{MATLAB} by solving \eqref{eq:cbf_selection} using \emph{YALMIP} \cite{Lofberg2004} with the 'fmincon option'. The calculation of $\mathfrak{b}(\boldsymbol{x},t)$ took $4.2$ seconds on an Intel Core i7-6600U with $16$ GB of RAM without maximizing $r$. In fact, increased oscillations in the control input were observed when we decided to maximize $r$.

\begin{figure}
\centering
\includegraphics[scale=0.395]{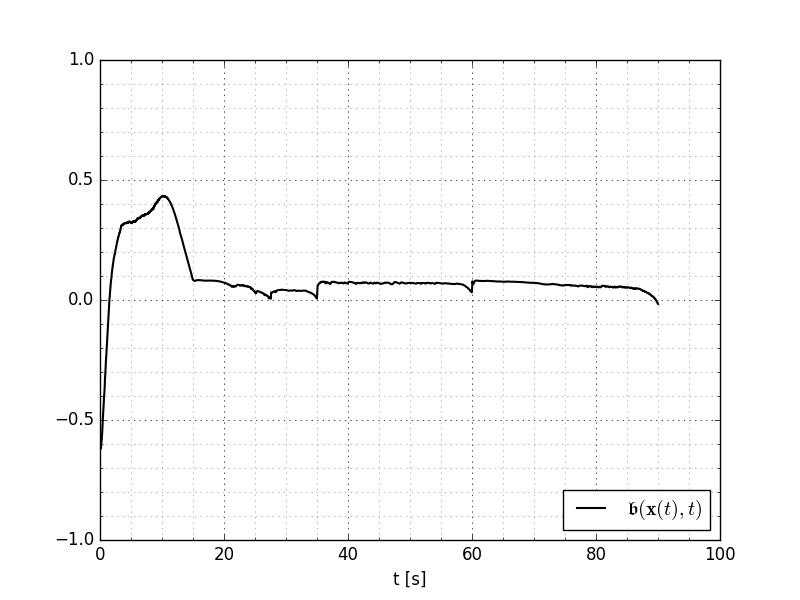}
\label{fig:scen2}
\caption{Barrier function evolution $\mathfrak{b}(\boldsymbol{x}(t),t)$.}
\label{fig:ex1}
\end{figure}

%\begin{figure*}
%\begin{subfigure}{0.48\textwidth}
%\hspace{-0.5cm}
%\includegraphics[scale=0.34]{figures/trajectory0_mod}
%\caption{Robot trajectories. The increasing color occupancy indicates the initial position and the evolution of the robots as time progresses.}
%\label{fig:21}
%\end{subfigure}
%\hspace{0.5cm}
%\begin{subfigure}{0.48\textwidth}
%\hspace{-0.5cm}
%\includegraphics[scale=0.47]{figures/input2}
%\caption{The $x$ and $y$ components $u_x$ and $u_y$ of robot $3$'s control input  $\boldsymbol{u}_3(\boldsymbol{x},t)$  plotted over time. Discontinuities are visible as expected.}
%\label{fig:22}
%\end{subfigure}
%\caption{Robot trajectories and control inputs.}
%\label{fig:ex2}
%\end{figure*}

\begin{figure}
\centering
\includegraphics[scale=0.295]{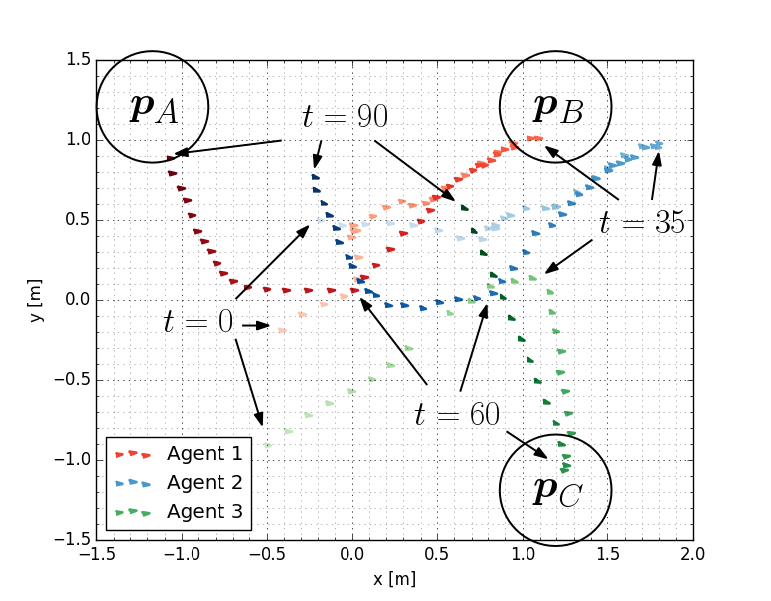}
\caption{Robot trajectories. The increasing color occupancy indicates  the evolution of the robots as time progresses.}
\label{fig:21}
\end{figure}

\begin{figure}
\centering
\includegraphics[scale=0.395]{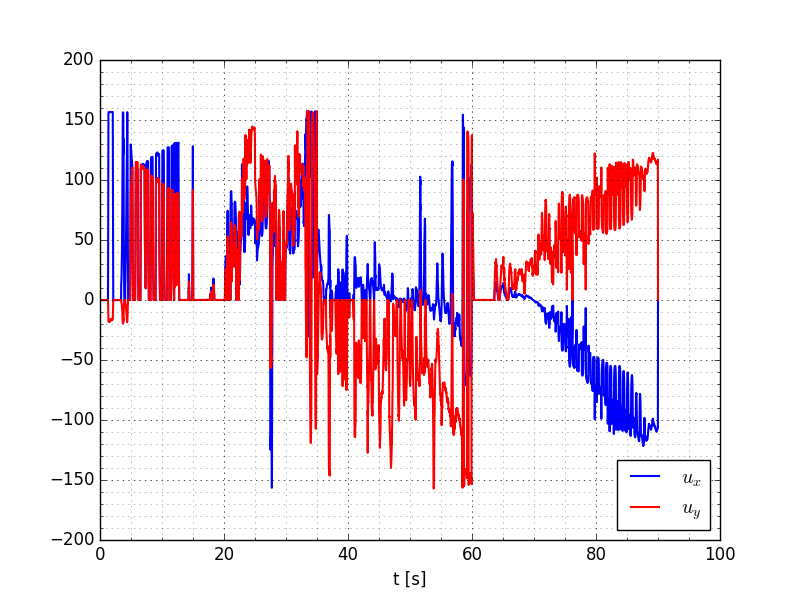}
\caption{The $x$ and $y$ components $u_x$ and $u_y$ of  $\boldsymbol{u}_3(\boldsymbol{x},t)$  plotted over time. Discontinuities and chattering are visible.}
\label{fig:22}
\end{figure}

The experimental result is shown in Figs. \ref{fig:ex1}-\ref{fig:22} as well as in \cite{experiment} where we provide a video of the experiment. To illustrate Remark \ref{rem:attrac}, we have intentionally chosen an initial condition $\boldsymbol{x}(0)$ of the robots that does not coincide with the initial condition $\boldsymbol{x}(0):=\boldsymbol{0}$ used in \eqref{eq:cbf_selection} to construct $\mathfrak{b}(\boldsymbol{x},t)$. In Fig. \ref{fig:ex1}, it is hence visible that initially $\mathfrak{b}(\boldsymbol{x}(0),0)\approx -0.62$. However, after approximately $t\approx 2$ sec, it holds that $\mathfrak{b}(\boldsymbol{x}(t),t)\ge 0$ and the robots have recovered from this situation. This is, in particular, a strength compared to our previous approach \cite{lindemann2018decentralized} where the control law would have not been defined in case of such a mismatch. Furthemore, Fig. \ref{fig:ex1} shows that $\mathfrak{b}(\boldsymbol{x}(t),t)\ge 0$ for the rest of the experiment so that it can be concluded that $(\boldsymbol{x},0)\models \phi$ or, to be more precise, $\rho^\phi(\boldsymbol{x},0)\ge r$ where $r=0.05$ was obtained by the solution of \eqref{eq:cbf_selection}. Fig. \ref{fig:21} shows the corresponding robot trajectories. As emphasized in Section \ref{sec:controller}, the control law $\boldsymbol{u}_i(\boldsymbol{x},t)$ is discontinuous. This is shown by plotting the $x$ and $y$ component of $\boldsymbol{u}_3(\boldsymbol{x},t)$ in Fig. \ref{fig:22}. We intentionally avoided to use an additional filter on  $\boldsymbol{u}_i(\boldsymbol{x},t)$ to smoothen the control input in order to show the nature of the discontinuous control law. The low-level PID controllers, however, filter $\boldsymbol{u}_i(\boldsymbol{x},t)$ when applied to the motors of the robots. We further remark that using linear functions $\gamma_l(t)$ to construct $\mathfrak{b}(\boldsymbol{x},t)$ as introduced in Section \ref{sec:construction_rules} compared to exponential ones as presented in \cite{lindemann2019decentralized} is beneficial since input saturations are less likely to occur. An exponential function $\gamma_l(t)$ would, for some $t$, induce high control inputs, while for other $t$ nearly no control action would be needed. A linear function $\gamma_l(t)$ distributes the needed control action more uniformly over time and is hence more suited for experiments. Finally, note that collisions are avoided by the use of $f_i^\text{u}(\boldsymbol{x},t)$, especially in the first $5$ seconds where a collision would occur between robot $1$ and $2$ without induced dynamical couplings in $f_i^\text{u}(\boldsymbol{x},t)$. We remark that approaches such as \cite{raman1} are not applicable here. First of all, the induced computational complexity does not allow to obtain the solution to a mixed linear program in reasonable time; \cite{raman1} also does not allow for nonlinear predicate functions as required by $\phi$. Existing approaches work with discrete-time systems. We, however, directly consider continuous-time systems and provide continuous-time satisfaction guarantees.

\section{Conclusion}
\label{sec:conclusion}
 
We have proposed a collaborative feedback control strategy for dynamically coupled multi-agent systems under a set of signal temporal logic tasks. For each agent, we have first derived a collaborative decentralized feedback control law that guarantees the satisfaction of all tasks. This control law is discontinuous, hence  Filippov solutions and  nonsmooth analysis is used, and based on the existence of a control barrier function that accounts for the semantics of the signal temporal logic task at hand. We have then presented how a control barrier function can be constructed for a fragment of signal temporal logic tasks by solving an optimization problem. Finally, we have validated our theoretical results in an experiment including  three omnidirectional robots.

\appendix[Proof of Theorem \ref{theorem1}]
We will show, in three parts, that \eqref{eq:qp_conv} is always feasible, that there exist Fillipov solutions to \eqref{system_noise} under $\boldsymbol{u}_i(\bar{\boldsymbol{x}}_k,t)$, and that Corollary \ref{corollary_phi_sat} can be applied. According to the assumptions, $\bar{\mathfrak{b}}_k(\bar{\boldsymbol{x}}_k,t)$ is a cCBF for each time interval $[s^k_j,s^k_{j+1})$ and $\bar{\mathfrak{b}}_k(\bar{\boldsymbol{x}}_k,t)$ is again piecewise continuous in $t$. As argued in the proof of Corollary \ref{corollary_phi_sat}, it is hence sufficient to look at each time interval $[s^k_j,s^k_{j+1})$ separately. Next, define 
\begin{align*}
\mathfrak{B}_{j,i}^k&:=\Big\{(\bar{\boldsymbol{x}}_k,t)\in\mathfrak{D}_k\times(s^k_j,s^k_{j+1})|\frac{\partial \bar{\mathfrak{b}}_k(\bar{\boldsymbol{x}}_k,t)}{\partial \boldsymbol{x}_i}= \boldsymbol{0}\Big\}\setminus \mathfrak{B}_{j}^k\\ 
\bar{\mathfrak{B}}_{j,i}^k&:=\Big\{(\bar{\boldsymbol{x}}_k,t)\in\mathfrak{D}_k\times(s^k_j,s^k_{j+1})|\frac{\partial \bar{\mathfrak{b}}_k(\bar{\boldsymbol{x}}_k,t)}{\partial \boldsymbol{x}_i}\neq \boldsymbol{0}\Big\}
\end{align*}

We remark that $\mathfrak{B}_{j}^k\cup\mathfrak{B}_{j,i}^k\cup\bar{\mathfrak{B}}_{j,i}^k=\mathfrak{D}_k\times(s_j^k,s_{j+1}^k)$ and that $\mathfrak{B}_{j}^k$, $\mathfrak{B}_{j,i}^k$, and $\bar{\mathfrak{B}}_{j,i}^k$ are disjoint sets. To understand the details of the following proof, note that $\mathfrak{B}_{j}^k$ and $\mathfrak{B}_{j,i}^k$ can not be closed sets (note that $(s_j^k,s_{j+1}^k)$ is open) and that information regarding these sets being open or not is not available. We will, however, show and use the fact that $\bar{\mathfrak{B}}_{j,i}^k$ is open.

\emph{Part 1 - Feasibility of  \eqref{eq:qp_conv}:} We next show that \eqref{eq:qp_conv} is always feasible and distinguish between three cases indicated by $\mathfrak{B}_{j}^k$, $\mathfrak{B}_{j,i}^k$, and $\bar{\mathfrak{B}}_{j,i}^k$. It will turn out that $\boldsymbol{u}_i(\bar{\boldsymbol{x}}_k,t)$  may be discontinuous on the boundaries of $\mathfrak{B}_{j}^k$, $\mathfrak{B}_{j,i}^k$, and $\bar{\mathfrak{B}}_{j,i}^k$.

Case 1 applies when $(\bar{\boldsymbol{x}}_k,t)\in \mathfrak{B}_{j}^k$. This is equivalent to $(\bar{\boldsymbol{x}}_k,t)\in\mathfrak{D}_k\times(s_j^k,s_{j+1}^k)$ such that $\frac{\partial \bar{\mathfrak{b}}_k(\bar{\boldsymbol{x}}_k,t)}{\partial \bar{\boldsymbol{x}}_k}=\boldsymbol{0}$ (which is equivalent to \re{$\sum_{v\in\mathcal{V}_k}\|\frac{\partial \bar{\mathfrak{b}}_k(\bar{\boldsymbol{x}}_k,t)}{\partial \boldsymbol{x}_v}\|_1= 0$}) and implies $\frac{\partial \bar{\mathfrak{b}}_k(\bar{\boldsymbol{x}}_k,t)}{\partial \boldsymbol{x}_i}=\boldsymbol{0}$; \eqref{eq:const_qp} reduces to $\omega_k(\bar{\boldsymbol{x}}_k,t)  \ge 0$ since $D_i(\bar{\boldsymbol{x}}_k,t)=1$ so that \eqref{eq:const_qp} is satisfied due to Assumption \ref{ass3}. Hence $\boldsymbol{u}_i(\bar{\boldsymbol{x}}_k,t)=\boldsymbol{0}$ is the optimal solution to \eqref{eq:qp_conv}. 

Case 2 applies when $(\bar{\boldsymbol{x}}_k,t)\in \mathfrak{B}_{j,i}^k$. This is equivalent to $(\bar{\boldsymbol{x}}_k,t)\in\mathfrak{D}_k\times(s_j^k,s_{j+1}^k)$ such that $\frac{\partial \bar{\mathfrak{b}}_k(\bar{\boldsymbol{x}}_k,t)}{\partial \bar{\boldsymbol{x}}_k}\neq \boldsymbol{0}$ (which is equivalent to \re{$\sum_{v\in\mathcal{V}_k}\|\frac{\partial \bar{\mathfrak{b}}_k(\bar{\boldsymbol{x}}_k,t)}{\partial \boldsymbol{x}_v}\|_1\neq 0$}) and $\frac{\partial \bar{\mathfrak{b}}_k(\bar{\boldsymbol{x}}_k,t)}{\partial \boldsymbol{x}_i}=\boldsymbol{0}$. The optimal solution to \eqref{eq:qp_conv} is again $\boldsymbol{u}_i(\bar{\boldsymbol{x}}_k,t)=\boldsymbol{0}$ since \eqref{eq:const_qp} is trivially satisfied (note that $D_i(\bar{\boldsymbol{x}}_k,t)=0$). 

Case 3 applies when $(\bar{\boldsymbol{x}}_k,t)\in \bar{\mathfrak{B}}_{j,i}^k$. This is equivalent to $(\bar{\boldsymbol{x}}_k,t)\in\mathfrak{D}_k\times(s_j^k,s_{j+1}^k)$ such that   $\frac{\partial \bar{\mathfrak{b}}_k(\bar{\boldsymbol{x}}_k,t)}{\partial \boldsymbol{x}_i}\neq\boldsymbol{0}$ so that \eqref{eq:qp_conv} is feasible. Note again that $\frac{\partial \bar{\mathfrak{b}}_k(\bar{\boldsymbol{x}}_k,t)}{\partial \boldsymbol{x}_i}\neq\boldsymbol{0}$ implies $\frac{\partial \bar{\mathfrak{b}}_k(\bar{\boldsymbol{x}}_k,t)}{\partial \boldsymbol{x}_i}g_i(\boldsymbol{x}_i,t)\neq\boldsymbol{0}$; $\boldsymbol{u}_i(\bar{\boldsymbol{x}}_k,t)$ is locally Lipschitz continuous on $\text{int}(\bar{\mathfrak{B}}_{j,i}^k)$  where $\text{int}(\cdot)$ denotes the interior of a set. This follows by virtue of \cite[Thm. 3]{ames2017control} and since all functions in \eqref{eq:qp_conv}  are locally Lipschitz continuous on $\text{int}(\bar{\mathfrak{B}}_{j,i}^k)$. In particular, $D_i(\bar{\boldsymbol{x}}_k,t)$ is locally Lipschitz continuous on $\text{int}(\mathfrak{B}_{j,i}^k)\cup\text{int}(\bar{\mathfrak{B}}_{j,i}^k)$ and  $\frac{\partial \bar{\mathfrak{b}}_k(\bar{\boldsymbol{x}}_k,t)}{\partial \boldsymbol{x}_i}$, $\frac{\partial \bar{\mathfrak{b}}_k(\bar{\boldsymbol{x}}_k,t)}{\partial t}$, $f_i(\boldsymbol{x}_i,t)$, $g_i(\boldsymbol{x}_i,t)$, and $\alpha_k(\bar{\mathfrak{b}}_k(\bar{\boldsymbol{x}}_k,t))$ are locally Lipschitz continuous on $\text{int}(\mathfrak{B}_{j}^k)\cup\text{int}(\mathfrak{B}_{j,i}^k)\cup\text{int}(\bar{\mathfrak{B}}_{j,i}^k)$. 

The optimization problem \eqref{eq:qp_conv} is hence always feasible and $\boldsymbol{u}_i(\bar{\boldsymbol{x}}_k,t)$ is locally Lipschitz continuous on $\text{int}(\mathfrak{B}_{j}^k)$, $\text{int}(\mathfrak{B}_{j,i}^k)$, and $\text{int}(\bar{\mathfrak{B}}_{j,i}^k)$. 

\emph{Part 2 - Existence of Filippov Solutions to \eqref{system_noise} under $\boldsymbol{u}_i(\bar{\boldsymbol{x}}_k,t)$:} The control law $\boldsymbol{u}_i(\bar{\boldsymbol{x}}_k,t)$ may, as indicated above, be discontinuous; $\boldsymbol{u}_i(\bar{\boldsymbol{x}}_k,t)$ is, however, locally bounded and measurable on $\mathfrak{D}_k\times(s_j^k,s_{j+1}^k)$ as argued next. In particular, we already know that $\boldsymbol{u}_i(\bar{\boldsymbol{x}}_k,t)$ is locally bounded on $\text{int}(\mathfrak{B}_{j}^k)$, $\text{int}(\mathfrak{B}_{j,i}^k)$, and $\text{int}(\bar{\mathfrak{B}}_{j,i}^k)$ due to being locally Lipschitz continuous on these domains. If we ensure that $\boldsymbol{u}_i(\bar{\boldsymbol{x}}_k,t)$ is also locally bounded on the boundaries of $\mathfrak{B}_{j}^k$, $\mathfrak{B}_{j,i}^k$, and $\bar{\mathfrak{B}}_{j,i}^k$, we can conclude that $\boldsymbol{u}_i(\bar{\boldsymbol{x}}_k,t)$ is locally bounded on $\mathfrak{D}_k\times(s_j^k,s_{j+1}^k)$. Therefore, we next systematically investigate the cases where $(\bar{\boldsymbol{x}}_k,t)$ is in $\{\text{bd}(\mathfrak{B}_{j}^k)\cap\text{bd}(\mathfrak{B}_{j,i}^k)\}\setminus\text{bd}(\bar{\mathfrak{B}}_{j,i}^k)$ (Cases 1 or 2), $\{\text{bd}(\mathfrak{B}_{j}^k)\cap\text{bd}(\bar{\mathfrak{B}}_{j,i}^k)\}\setminus\text{bd}(\mathfrak{B}_{j,i}^k)$ (Cases 1 or 3), $\{\text{bd}(\mathfrak{B}_{j,i}^k)\cap\text{bd}(\bar{\mathfrak{B}}_{j,i}^k)\}\setminus \text{bd}(\mathfrak{B}_{j}^k)$ (Cases 2 or 3), and $\text{bd}(\mathfrak{B}_{j}^k)\cap\text{bd}(\mathfrak{B}_{j,i}^k)\cap\text{bd}(\bar{\mathfrak{B}}_{j,i}^k)$ (Cases 1, 2, or 3) where $\text{bd}(\cdot)$ denotes the boundary of a set. 

When $(\bar{\boldsymbol{x}}_k,t)\in\{\text{bd}(\mathfrak{B}_{j}^k)\cap\text{bd}(\mathfrak{B}_{j,i}^k)\}\setminus\text{bd}(\bar{\mathfrak{B}}_{j,i}^k)$, either $(\bar{\boldsymbol{x}}_k,t)\in\mathfrak{B}_{j}^k$ or $(\bar{\boldsymbol{x}}_k,t)\in\mathfrak{B}_{j,i}^k$. Either way, due to continuity (recall that $\boldsymbol{u}_i(\bar{\boldsymbol{x}}_i,t)=\boldsymbol{0}$ in Case 1 and 2) there exists a neighborhood $\mathcal{U}\subseteq \{\mathfrak{B}_{j}^k\cup\mathfrak{B}_{j,i}^k\}\setminus\bar{\mathfrak{B}}_{j,i}^k$ around $(\bar{\boldsymbol{x}}_k,t)$ so that, for each $(\bar{\boldsymbol{x}}_k',t')\in\mathcal{U}$, $\|\boldsymbol{u}_i(\bar{\boldsymbol{x}}_k',t')\|=\boldsymbol{0}$. Consequently, $\boldsymbol{u}_i(\bar{\boldsymbol{x}}_k,t)$ is locally bounded on $\{\text{bd}(\mathfrak{B}_{j}^k)\cap\text{bd}(\mathfrak{B}_{j,i}^k)\}\setminus\text{bd}(\bar{\mathfrak{B}}_{j,i}^k)$.  

When $(\bar{\boldsymbol{x}}_k,t)\in\{\text{bd}(\mathfrak{B}_{j}^k)\cap\text{bd}(\bar{\mathfrak{B}}_{j,i}^k)\}\setminus\text{bd}(\mathfrak{B}_{j,i}^k)$, either $(\bar{\boldsymbol{x}}_k,t)\in\mathfrak{B}_{j}^k$ or $(\bar{\boldsymbol{x}}_k,t)\in\bar{\mathfrak{B}}_{j,i}^k$. Note that $\omega_k(\bar{\boldsymbol{x}}_k,t) \ge \epsilon_k$ if $(\bar{\boldsymbol{x}}_k,t)\in\mathfrak{B}_{j}^k$ due to Assumption \ref{ass3}. Recall also that $\omega_k(\bar{\boldsymbol{x}}_k,t)$ is continuous on $\mathfrak{D}_k\times(s_j^k,s_{j+1}^k)$. By the definition of continuity it follows that for a given $\epsilon_k>0$ (in this case, the $\epsilon_k$ from Assumption \ref{ass3}) there exists a $\delta_k>0$ so that for each $(\bar{\boldsymbol{x}}_k',t')$ with $\|\begin{bmatrix} {\bar{\boldsymbol{x}}_k}^{'T} & t'\end{bmatrix}^T-\begin{bmatrix} {\bar{\boldsymbol{x}}_k}^T & t\end{bmatrix}^T\|<\delta_k$ it holds that $\omega_k(\bar{\boldsymbol{x}}_k,t)-\epsilon_k<\omega_k(\bar{\boldsymbol{x}}_k',t')<\omega_k(\bar{\boldsymbol{x}}_k,t)+\epsilon_k$ so that consequently $\omega_k(\bar{\boldsymbol{x}}_k',t')\ge 0$. Hence, there exists a neighborhood $\mathcal{U}\subseteq \{\mathfrak{B}_{j}^k\cup\bar{\mathfrak{B}}_{j,i}^k\}\setminus\mathfrak{B}_{j,i}^k$ around $(\bar{\boldsymbol{x}}_k,t)$ so that, for each $(\bar{\boldsymbol{x}}_k',t')\in\mathcal{U}$,  either $\|\boldsymbol{u}_i(\bar{\boldsymbol{x}}_k',t')\|=\boldsymbol{0}$ (if $(\bar{\boldsymbol{x}}_k',t')\in\mathfrak{B}_{j}^k\cap \mathcal{U}$) or $\omega_k(\bar{\boldsymbol{x}}_k',t')\ge 0$ (if $(\bar{\boldsymbol{x}}_k',t')\in\bar{\mathfrak{B}}_{j,i}^k\cap \mathcal{U}$). For the latter case, i.e., $(\bar{\boldsymbol{x}}_k',t')\in\bar{\mathfrak{B}}_{j,i}^k\cap \mathcal{U}$, note that a feasible (not necessarily optimal) and analytic control law for \eqref{eq:const_qp} is
\begin{align*}
\boldsymbol{u}_i^{\text{feas}}(\bar{\boldsymbol{x}}_k',t')&:={g_i(\boldsymbol{x}_i',t')}^TG_i(\boldsymbol{x}_i',t')^{-1}(-f_i(\boldsymbol{x}_i',t')+\boldsymbol{v}_i^{\text{feas}})
\end{align*}
where $\boldsymbol{x}_i'$ is the corresponding element in $\bar{\boldsymbol{x}}_k'$, $\boldsymbol{v}_i^{\text{feas}}$ is explained in the remainder, and where the inverse of $G_i(\boldsymbol{x}_i',t'):=g_i(\boldsymbol{x}_i',t'){g_i(\boldsymbol{x}_i',t')}^T$ exists due to Assumption \ref{ass1} so that 
\begin{align*}
\|\boldsymbol{u}_i^{\text{feas}}(\bar{\boldsymbol{x}}_k',t')\|\le C_{g_i}C_{G_i}(C_{f_i}+\|\boldsymbol{v}_i^{\text{feas}}\|)
\end{align*}
where $C_{g_i}$, $C_{G_i}$, and $C_{f_i}$ are upper bounds on $\|g_i(\boldsymbol{x}_i,t)\|$, $\|G_i(\boldsymbol{x}_i,t)\|$, and  $\|f_i(\boldsymbol{x}_i,t)\|$ that follow due to continuity of $g_i(\boldsymbol{x}_i,t)$, $G_i(\boldsymbol{x}_i,t)$, and  $f_i(\boldsymbol{x}_i,t)$ on the bounded domain $\mathfrak{D}_k$. Note especially that $G_i(\boldsymbol{x}_i,t)$ is upper bounded by the inverse of the smallest singular value of $G_i(\boldsymbol{x}_i,t)$ when the max matrix norm is used \cite[Ch.5.6]{horn1990matrix}. We next show how to select $\boldsymbol{v}_i^\text{feas}$ and that $\|\boldsymbol{v}_i^\text{feas}\|$ is also upper bounded. Using $\boldsymbol{u}_i^\text{feas}(\bar{\boldsymbol{x}}_k',t')$, \eqref{eq:const_qp} reduces to  
\begin{align}\label{eq:const_qp_reduced}
\begin{split}
&\frac{\partial \bar{\mathfrak{b}}_k(\bar{\boldsymbol{x}}_k',t')}{\partial \boldsymbol{x}_i}\boldsymbol{v}_i^\text{feas}\ge \re{\Big\|{\frac{\partial \bar{\mathfrak{b}}_k(\bar{\boldsymbol{x}}_k,t)}{\partial \boldsymbol{x}_i}}\Big\|_1 C}-D_i(\bar{\boldsymbol{x}}_k',t')\omega_k(\bar{\boldsymbol{x}}_k',t').
\end{split}
\end{align} 
We select \re{$\boldsymbol{v}_i^\text{feas}(\bar{\boldsymbol{x}}_k',t'):=\text{sgn}\big(\frac{\partial \bar{\mathfrak{b}}_k(\bar{\boldsymbol{x}}_k',t')}{\partial \boldsymbol{x}_i}\big)^T\kappa_i$ where $\text{sgn}(\cdot)$ is the element-wise sign operator} so that \eqref{eq:const_qp_reduced} becomes
\begin{align}\label{eq:const_qp_reduced2}
\begin{split}
&\re{\Big\|\frac{\partial \bar{\mathfrak{b}}_k(\bar{\boldsymbol{x}}_k',t')}{\partial \boldsymbol{x}_i}\Big\|_1\Big(\kappa_i -C+\frac{\omega_k(\bar{\boldsymbol{x}}_k',t')}{\sum_{v\in\mathcal{V}_k}\| \frac{\partial \bar{\mathfrak{b}}_k(\bar{\boldsymbol{x}}_k',t')}{\partial \boldsymbol{x}_v}\|_1}\Big) \ge 0.}
\end{split}
\end{align}
In particular, it holds that \eqref{eq:const_qp_reduced2} is satisfied if \re{$\kappa_i:=C$} (recall that $\omega_k(\bar{\boldsymbol{x}}_k',t')\ge 0$ if $(\bar{\boldsymbol{x}}_k',t')\in\bar{\mathfrak{B}}_{j,i}^k\cap \mathcal{U}$) \re{so that $\|\boldsymbol{v}_i^\text{feas}(\bar{\boldsymbol{x}}_k',t')\|\le C$}. Consequently, $\|\boldsymbol{u}_i(\bar{\boldsymbol{x}}_k',t')\|\le \|\boldsymbol{u}_i^\text{feas}(\bar{\boldsymbol{x}}_k',t')\|\le C_{g_i}C_{G_i}(C_{f_i}+\re{C})$ and $\boldsymbol{u}_i(\bar{\boldsymbol{x}}_k,t)$ is locally bounded on $\{\text{bd}(\mathfrak{B}_{j}^k)\cap\text{bd}(\bar{\mathfrak{B}}_{j,i}^k)\}\setminus\text{bd}(\mathfrak{B}_{j,i}^k)$. 

When $(\bar{\boldsymbol{x}}_k,t)\in\{\text{bd}(\mathfrak{B}_{j,i}^k)\cap\text{bd}(\bar{\mathfrak{B}}_{j,i}^k)\}\setminus\text{bd}(\mathfrak{B}_{j}^k)$, either $(\bar{\boldsymbol{x}}_k,t)\in\mathfrak{B}_{j,i}^k$ or $(\bar{\boldsymbol{x}}_k,t)\in\bar{\mathfrak{B}}_{j,i}^k$ and a similar analysis can be made as above. In particular, then there exists a neighborhood $\mathcal{U}\subseteq \{\mathfrak{B}_{j,i}^k\cup\bar{\mathfrak{B}}_{j,i}^k\}\setminus \mathfrak{B}_{j}^k$ around $(\bar{\boldsymbol{x}}_k,t)$ so that, for each $(\bar{\boldsymbol{x}}_k',t')\in\mathcal{U}$, either $\|\boldsymbol{u}_i(\bar{\boldsymbol{x}}_k',t')\|=\boldsymbol{0}$ (if $(\bar{\boldsymbol{x}}_k',t')\in\mathfrak{B}_{j,i}^k\cap \mathcal{U}$) or $\sum_{v\in\mathcal{V}_k}\re{\| \frac{\partial \bar{\mathfrak{b}}_k(\bar{\boldsymbol{x}}_k',t')}{\partial \boldsymbol{x}_v}\|_1}\ge \nu$ for some $\nu>0$ (if $(\bar{\boldsymbol{x}}_k',t')\in\bar{\mathfrak{B}}_{j,i}^k\cap \mathcal{U}$) since $(\bar{\boldsymbol{x}}_k',t')\notin{\mathfrak{B}}_{j}^k$ and again due to continuity. In the latter case, selecting $\kappa_i:=\re{C}-\frac{\omega_k(\bar{\boldsymbol{x}}_k',t')}{\nu}$ satisfies \eqref{eq:const_qp_reduced2}. The same arguments as before then show that $\boldsymbol{u}_i(\bar{\boldsymbol{x}}_k,t)$ is locally bounded on $\{\text{bd}(\mathfrak{B}_{j,i}^k)\cap\text{bd}(\bar{\mathfrak{B}}_{j,i}^k)\}\setminus\text{bd}(\mathfrak{B}_{j}^k)$. 

When  $(\bar{\boldsymbol{x}}_k,t)\in\text{bd}(\mathfrak{B}_{j}^k)\cap\text{bd}(\mathfrak{B}_{j}^k)\cap\text{bd}(\bar{\mathfrak{B}}_{j,i}^k)$, it can again be shown that $\boldsymbol{u}_i(\bar{\boldsymbol{x}}_k,t)$ is locally bounded on $\text{bd}(\mathfrak{B}_{j}^k)\cap\text{bd}(\mathfrak{B}_{j,i}^k)\cap\text{bd}(\bar{\mathfrak{B}}_{j,i}^k)$. The proof is straighforward using the same arguments as in the previous discussion and omitted. 

It follows that $\boldsymbol{u}_i(\bar{\boldsymbol{x}}_k,t)$ is locally bounded on $\text{bd}(\mathfrak{B}_{j}^k)$, $\text{bd}(\mathfrak{B}_{j,i}^k)$, and $\text{bd}(\bar{\mathfrak{B}}_{j,i}^k)$. Since we have already concluded that the same holds on $\text{int}(\mathfrak{B}_{j}^k)$, $\text{int}(\mathfrak{B}_{j,i}^k)$, and $\text{int}(\bar{\mathfrak{B}}_{j,i}^k)$, $\boldsymbol{u}_i(\bar{\boldsymbol{x}}_k,t)$ is consequently locally bounded  on $\mathfrak{B}_{j}^k\cup\mathfrak{B}_{j,i}^k\cup\bar{\mathfrak{B}}_{j,i}^k=\mathfrak{D}_k\times(s_j^k,s_{j+1}^k)$. To see that $\boldsymbol{u}_i(\bar{\boldsymbol{x}}_k,t)$ is measureable, note that $\mathfrak{B}_j^k$, $\mathfrak{B}_{j,i}^k$, and $\bar{\mathfrak{B}}_{j,i}^k$ are measurable sets. The product of measurable functions is measurable and the indicator function (here used to indicate Cases 1, 2, and 3) defined on measurable sets is measurable so that $\boldsymbol{u}_i(\bar{\boldsymbol{x}}_k,t)$ is measurable. Consequently, the multi-agent system described by the stacked dynamics of each agent in \eqref{system_noise} admits Filippov solutions $\boldsymbol{x}:[t_0,t_1]\to\mathbb{R}^n$ from each initial condition in $\mathfrak{D} \times\mathbb{R}_{\ge 0}$ where $\mathfrak{D}:=\mathfrak{D}_{j_1}\times\hdots\times\mathfrak{D}_{j_{|{\mathcal{V}}_k|}}$ for $j_1,\hdots,j_{|{\mathcal{V}}_k|}\in{\mathcal{V}}_k$. 

\emph{Part 3 - Application of Corollary \ref{corollary_phi_sat}:} For each  $\bar{\mathfrak{b}}_k(\bar{\boldsymbol{x}}_k,t)$, the individual solutions $\boldsymbol{u}_i(\bar{\boldsymbol{x}}_k,t)$ to \eqref{eq:qp_conv} result in 
\begin{align}
&\hspace{-0.35cm}\sum_{i\in\mathcal{V}_k}\frac{\partial \bar{\mathfrak{b}}_k(\bar{\boldsymbol{x}}_k,t)}{\partial \boldsymbol{x}_i}(f_i(\boldsymbol{x}_i,t)+g_i(\boldsymbol{x}_i,t)\boldsymbol{u}_i(\bar{\boldsymbol{x}}_k,t))\ge \nonumber\\
 &\hspace{0.8cm}\sum_{i\in\mathcal{V}_k}\Big(-D_i(\bar{\boldsymbol{x}}_k,t)\omega_k(\bar{\boldsymbol{x}}_k,t)+\re{\Big\|{\frac{\partial \bar{\mathfrak{b}}_k(\bar{\boldsymbol{x}}_k,t)}{\partial \boldsymbol{x}_i}}\Big\|_1}C\Big)\nonumber\\
 \Leftrightarrow &\;\frac{\partial \bar{\mathfrak{b}}_k(\bar{\boldsymbol{x}}_k,t)}{\partial \bar{\boldsymbol{x}}_k}(\bar{f}_k(\bar{\boldsymbol{x}}_k,t)+\bar{g}_k(\bar{\boldsymbol{x}}_k,t)\bar{\boldsymbol{u}}_k(\bar{\boldsymbol{x}}_k,t)) \ge \nonumber\\
 & \hspace{1.5cm}-\omega_k(\bar{\boldsymbol{x}}_k,t)+\sum_{i\in\mathcal{V}_k}\re{\Big\|{\frac{\partial \bar{\mathfrak{b}}_k(\bar{\boldsymbol{x}}_k,t)}{\partial \boldsymbol{x}_i}}\Big\|_1}C\nonumber\\
 \begin{split}\label{eq:result_barrier}
 \re{\Leftrightarrow} &\;\frac{\partial \bar{\mathfrak{b}}_k(\bar{\boldsymbol{x}}_k,t)}{\partial \bar{\boldsymbol{x}}_k}(\bar{f}_k(\bar{\boldsymbol{x}}_k,t)+\bar{g}_k(\bar{\boldsymbol{x}}_k,t)\bar{\boldsymbol{u}}_k(\bar{\boldsymbol{x}}_k,t))\ge \\
 & \hspace{1.9cm}-\omega_k(\bar{\boldsymbol{x}}_k,t)+\re{\Big\|{\frac{\partial \bar{\mathfrak{b}}_k(\bar{\boldsymbol{x}}_k,t)}{\partial \bar{\boldsymbol{x}}_k}}\Big\|_1}C
 \end{split}
\end{align}
\re{where the last equivalence follows by the definition of the sum norm.}

In our analysis below, it is crucial to note that $\bar{\mathfrak{B}}_{j,i}^k$ is open, which we show next. Denote by $\text{inv}\Big(\frac{\partial \bar{\mathfrak{b}}_k(\bar{\boldsymbol{x}}_k,t)}{\partial \boldsymbol{x}_i}(\mathcal{O}^{n_i})\Big)$ the inverse image of $\frac{\partial \bar{\mathfrak{b}}_k(\bar{\boldsymbol{x}}_k,t)}{\partial \boldsymbol{x}_i}$ under $\mathcal{O}^{n_i}$ where $\mathcal{O}:=(-\infty,0)\cup(0,\infty)$. Now, $\text{inv}\Big(\frac{\partial \bar{\mathfrak{b}}_k(\bar{\boldsymbol{x}}_k,t)}{\partial \boldsymbol{x}_i}(\mathcal{O}^{n_i})\Big)$ is open since $\mathcal{O}^{n_i}$ is open and since the inverse image of a continuous function under an open set is open \cite[Prop. 1.4.4]{aubin2009set}. It then holds that
\begin{align*}
\bar{\mathfrak{B}}_{j,i}^k=\{\mathfrak{D}_k\times(s^k_j,s^k_{j+1})\}\cap \text{inv}\Big(\frac{\partial \bar{\mathfrak{b}}_k(\bar{\boldsymbol{x}}_k,t)}{\partial \boldsymbol{x}_i}(\mathcal{O}^{n_i})\Big)
\end{align*} 
 is open since the intersection of open sets is open.

We next show that \eqref{eq:result_barrier} implies \eqref{eq:barrier_ineq} so that Corollary \ref{corollary_phi_sat} can be applied for each $\bar{\mathfrak{b}}_k(\bar{\boldsymbol{x}}_k,t)$. Since $\bar{f}_k(\bar{\boldsymbol{x}}_k,t)$ is locally Lipschitz continuous, it follows that $\hat{\mathcal{L}}_{F[\bar{f}_k]}\bar{\mathfrak{b}}_k(\bar{\boldsymbol{x}}_k,t):=\big\{\frac{\partial \bar{\mathfrak{b}}_k(\bar{\boldsymbol{x}}_k,t)}{\partial \bar{\boldsymbol{x}}_k}\bar{f}_k(\bar{\boldsymbol{x}}_k,t)\big\}$. For $\hat{\mathcal{L}}_{F[\bar{g}_k\bar{\boldsymbol{u}}_k]}\bar{\mathfrak{b}}_k(\bar{\boldsymbol{x}}_k,t)$, we have to distinguish between the aforementioned three cases. First note that, if for each $i\in\mathcal{V}_k$ we have $(\bar{\boldsymbol{x}}_k,t)\in\bar{\mathfrak{B}}_{j,i}^k$ (Case 3), then $\hat{\mathcal{L}}_{F[\bar{g}_k\bar{\boldsymbol{u}}_k]}\bar{\mathfrak{b}}_k(\bar{\boldsymbol{x}}_k,t)=\big\{\frac{\partial \bar{\mathfrak{b}}_k(\bar{\boldsymbol{x}}_k,t)}{\partial \bar{\boldsymbol{x}}_k}\bar{g}_k(\bar{\boldsymbol{x}}_k,t)\bar{\boldsymbol{u}}_k(\bar{\boldsymbol{x}}_k,t)\big\}$. This in particular follows since $\bar{\mathfrak{B}}_{j,i}^k$ is open so that $\boldsymbol{u}_i(\bar{\boldsymbol{x}}_k,t)$ as well as $\bar{g}_k(\bar{\boldsymbol{x}}_k,t)$ are locally Lipschitz continuous on $\text{int}(\bar{\mathfrak{B}}_{j,i}^k)=\bar{\mathfrak{B}}_{j}^k$.  If for each $i\in\mathcal{V}_k$ we have $(\bar{\boldsymbol{x}}_k,t)\in\mathfrak{B}_{j,i}^k$ (Case 2), then $\hat{\mathcal{L}}_{F[\bar{g}_k\bar{\boldsymbol{u}}_k]}\bar{\mathfrak{b}}_k(\bar{\boldsymbol{x}}_k,t)= \{0\}$ since $\frac{\partial \bar{\mathfrak{b}}_k(\bar{\boldsymbol{x}}_k,t)}{\partial \boldsymbol{x}_i}=\boldsymbol{0}$ for each $i\in\mathcal{V}_k$. If for some agents $(\bar{\boldsymbol{x}}_k,t)\in\bar{\mathfrak{B}}_{j,i}^k$ while for others  $(\bar{\boldsymbol{x}}_k,t)\in\mathfrak{B}_{j,i}^k$ (i.e., a mix of Case 2 and Case 3), the resulting $\hat{\mathcal{L}}_{F[\bar{g}_k\bar{\boldsymbol{u}}_k]}\bar{\mathfrak{b}}_k(\bar{\boldsymbol{x}}_k,t)$ will still be a singleton. If we have $(\bar{\boldsymbol{x}}_k,t)\in\mathfrak{B}_{j}^k$ (Case 1), then $\hat{\mathcal{L}}_{F[\bar{g}_k\bar{\boldsymbol{u}}_k]}\bar{\mathfrak{b}}_k(\bar{\boldsymbol{x}}_k,t)=\{0\}$ since $\frac{\partial \bar{\mathfrak{b}}_k(\bar{\boldsymbol{x}}_k,t)}{\partial \bar{\boldsymbol{x}}_k}=\boldsymbol{0}$. Note that $\min \mathcal{S}=\mathcal{S}$ when $\mathcal{S}$ is a singleton and recall  $\omega_k(\bar{\boldsymbol{x}}_k,t)$ in \eqref{eq:omega}. Since $\hat{\mathcal{L}}_{F[\bar{f}_k]}\bar{\mathfrak{b}}_k(\bar{\boldsymbol{x}}_k,t)$, $\hat{\mathcal{L}}_{F[\bar{g}_k\bar{\boldsymbol{u}}_k]}\bar{\mathfrak{b}}_k(\bar{\boldsymbol{x}}_k,t)$, and $\frac{\partial \bar{\mathfrak{b}}_k(\bar{\boldsymbol{x}}_k,t)}{\partial t}$  are singletons,  \eqref{eq:result_barrier} is equivalent to
\begin{align}
\begin{split}\label{eq:barrier_result}
&\hspace{-0.15cm}\min \Big\{ \hat{\mathcal{L}}_{F[\bar{f}_k]}\bar{\mathfrak{b}}_k(\bar{\boldsymbol{x}}_k,t) \oplus\hat{\mathcal{L}}_{F[\bar{g}_k\bar{\boldsymbol{u}}_k]}\bar{\mathfrak{b}}_k(\bar{\boldsymbol{x}}_k,t) \oplus\big\{\frac{\partial \bar{\mathfrak{b}}_k(\bar{\boldsymbol{x}}_k,t)}{\partial t}\big\} \Big\}\\
 &\hspace{1cm}  \ge-\alpha_k(\bar{\mathfrak{b}}_k(\bar{\boldsymbol{x}}_k,t))+\re{\Big\|{\frac{\partial \bar{\mathfrak{b}}_k(\bar{\boldsymbol{x}}_k,t)}{\partial \bar{\boldsymbol{x}}_k}}\Big\|_1C}.
 \end{split}
\end{align}
Due to Lemma \ref{lemma_liederivate}, $\mathcal{L}_{F[\bar{f}_k+\bar{g}_k\bar{\boldsymbol{u}}_k]}\bar{\mathfrak{b}}_k(\bar{\boldsymbol{x}}_k,t)\subseteq \hat{\mathcal{L}}_{F[\bar{f}_k]}\bar{\mathfrak{b}}_k(\bar{\boldsymbol{x}}_k,t)\oplus\hat{\mathcal{L}}_{F[\bar{g}_k\bar{\boldsymbol{u}}_k]}\bar{\mathfrak{b}}_k(\bar{\boldsymbol{x}}_k,t)\oplus\big\{\frac{\partial \bar{\mathfrak{b}}_k(\bar{\boldsymbol{x}}_k,t)}{\partial t}\big\}$ so that $\min\mathcal{L}_{F[\bar{f}_k+\bar{g}_k\bar{\boldsymbol{u}}_k]}\bar{\mathfrak{b}}_k(\bar{\boldsymbol{x}}_k,t)\ge \min\big\{\hat{\mathcal{L}}_{F[\bar{f}_k]}\bar{\mathfrak{b}}_k(\bar{\boldsymbol{x}}_k,t)\oplus\hat{\mathcal{L}}_{F[\bar{g}_k\bar{\boldsymbol{u}}_k]}\bar{\mathfrak{b}}_k(\bar{\boldsymbol{x}}_k,t)\oplus\big\{\frac{\partial \bar{\mathfrak{b}}_k(\bar{\boldsymbol{x}}_k,t)}{\partial t}\big\}\big\}$. Consequently,  \eqref{eq:barrier_result} implies \eqref{eq:barrier_ineq} and $(\boldsymbol{x},0)\models \phi_1 \wedge \hdots\wedge \phi_K$ follows  by Corollary \ref{corollary_phi_sat}. 

%% you can choose not to have a title for an appendix
%% if you want by leaving the argument blank
%\section{}

%\section*{Acknowledgment}

\ifCLASSOPTIONcaptionsoff
  \newpage
\fi

% references section%

\bibliographystyle{IEEEtran}
\bibliography{literature}

%\begin{thebibliography}{1}
%\end{thebibliography}


 \begin{IEEEbiography}[{\includegraphics[width=1in,height=1.25in,clip,keepaspectratio]{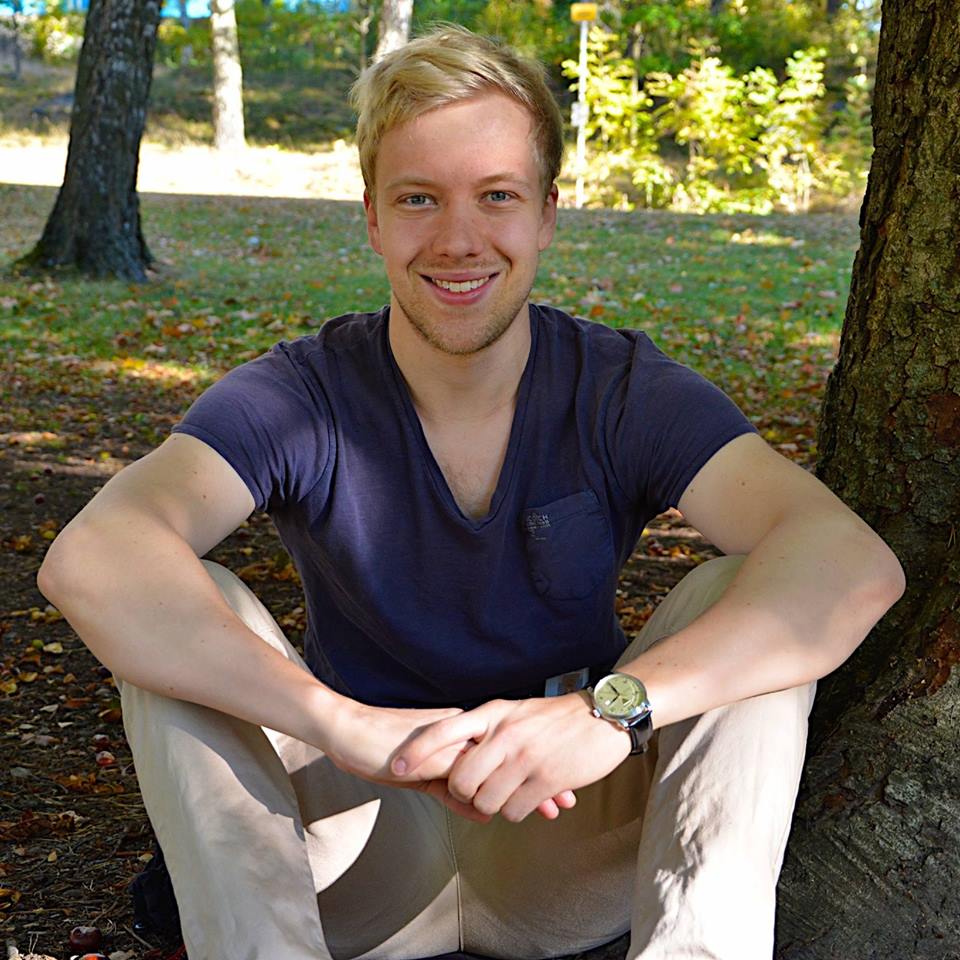}}]{Lars Lindemann} (S'17) was born in L\"ubbecke, Germany, in 1989. He received the B.Sc. degree in Electrical and Information Engineering and the B.Sc. degree in Engineering Management both from the Christian-Albrechts-University (CAU), Kiel, Germany, in 2014 and the M.Sc. degree in Systems, Control and Robotics from KTH Royal Institute of Technology, Stockholm, Sweden, in 2016.  Since June 2016, he is pursuing the Ph.D. degree at KTH Royal Institute of Technology, Stockholm, Sweden. His current research interests include control theory, formal methods, multi-agent systems, and autonomous systems. He was a  Best Student Paper Award Finalist at the 2018 American Control Conference and is a recipient of the Outstanding Student Paper Award of the  58th IEEE Conference on Decision and Control.

\end{IEEEbiography}

\begin{IEEEbiography}[{\includegraphics[width=1in,height=1.25in,clip,keepaspectratio]{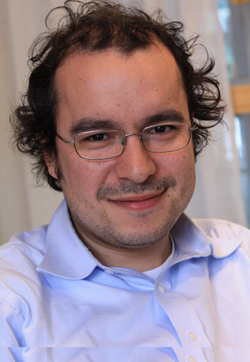}}]{Dimos V. Dimarogonas}
(M'10, SM'17) was born in Athens, Greece, in 1978. He received the Diploma in Electrical and Computer Engineering in 2001 and the Ph.D. in Mechanical Engineering in 2007, both from National Technical University of Athens (NTUA), Greece. Between 2007 and 2010, he held postdoctoral positions at the KTH Royal Institute of Technology, Dept of Automatic Control and MIT, Laboratory for Information and Decision Systems (LIDS). He is currently Professor at the Division of Decision and  Control Systems, School of Electrical Engineering and Computer Science, at KTH. His current research interests include multi-agent systems, hybrid systems and control, robot navigation and manipulation, human-robot-interaction and networked control. He serves in the Editorial Board of Automatica and the IEEE Transactions on Control of Network Systems and is a Senior Member of IEEE. He is a recipient of the ERC Starting Grant in 2014, the ERC Consolidator Grant in 2019, and the Knut och Alice Wallenberg Academy Fellowship in 2015.

\end{IEEEbiography}


%\vfill


\end{document}